\documentclass[journal]{IEEEtran}
\usepackage{amsmath}
\usepackage{amssymb}

\usepackage{graphicx}
\usepackage{epstopdf}
\usepackage{amsmath}
\usepackage{array}
\newcommand{\PreserveBackslash}[1]{\let\temp=\\#1\let\\=\temp}
\newcolumntype{C}[1]{>{\PreserveBackslash\centering}p{#1}}
\newcolumntype{R}[1]{>{\PreserveBackslash\raggedleft}p{#1}}
\newcolumntype{L}[1]{>{\PreserveBackslash\raggedright}p{#1}}
\usepackage[usenames]{color}
\usepackage{colortbl,booktabs}
\usepackage{tabularx}
\usepackage{multicol}
\usepackage{booktabs}
\usepackage[Symbol]{upgreek}
\usepackage{bm}
\usepackage{cite}
\usepackage{balance}
\usepackage{mathrsfs}
\usepackage{url}
\usepackage{threeparttable}
\usepackage{amsmath}
\usepackage{dcolumn}
\usepackage{multirow}
\usepackage{booktabs}
\usepackage{amsfonts}
\usepackage{enumerate}
\usepackage{subfigure}
\usepackage{algorithm}
\usepackage{algorithmic}
\begin{document}
	\title{Joint Hybrid and Passive RIS-Assisted Beamforming for  MmWave MIMO Systems Relying on Dynamically Configured Subarrays}
	\author{Chenghao Feng, Wenqian Shen, Jianping An, and Lajos Hanzo,~\IEEEmembership{Fellow,~IEEE}.
\thanks{This work was supported in part by the National Natural Science Foundation of China (NSFC) under Grant 61901034, and in part by the Open Research Fund of the Shaanxi Province Key Laboratory of Information Communication Network and Security under Grant ICNS201905.
                L. Hanzo would like to acknowledge the financial support of the Engineering and Physical Sciences Research Council projects EP/P034284/1 and EP/P003990/1 (COALESCE) as well as of the European Research Council's Advanced Fellow Grant QuantCom (Grant No. 789028) (\it{Corresponding author: Wenqian Shen.})}
        \thanks{
		C. Feng, W. Shen, and J. An are with the School of Information and Electronics, Beijing Institute of Technology, Beijing 100081, China (e-mails: cfeng@bit.edu.cn, shenwq@bit.edu.cn, and an@bit.edu.cn).\par
        L. Hanzo is with the Department of Electronics and Computer Science, University of Southampton, Southampton SO17 1BJ, UK (e-mail: lh@ecs.soton.ac.uk).
		}
	}{ \vspace{-10mm}}
	\maketitle{ \vspace{-15mm}}

	\begin{abstract}
    Reconfigurable intelligent surface (RIS) assisted millimeter-wave (mmWave) communication systems relying on hybrid beamforming structures are capable of achieving high spectral efficiency at a low hardware complexity and low power consumption.
    In this paper, we propose an RIS-assisted mmWave point-to-point system relying on dynamically configured sub-array connected hybrid beamforming structures.
    More explicitly, an energy-efficient analog beamformer relying on twin-resolution phase shifters is proposed.
    Then, we conceive a successive interference cancelation (SIC) based method for jointly designing the hybrid beamforming matrix of the base station (BS) and the passive beamforming matrix of the RIS.
    Specifically, the associated bandwidth-efficiency maximization problem is transformed into a series of sub-problems, where the sub-array of phase shifters and RIS elements are jointly optimized for maximizing each sub-array's rate.
    Furthermore, a greedy method is proposed for determining the phase shifter configuration of each sub-array.
    We then propose to update the RIS elements relying on a complex circle manifold (CCM)-based method.
    %A complex circle manifold (CCM)-based method is exploited for maximizing the upper bound of bandwidth efficiency.
    The proposed dynamic sub-connected structure as well as the proposed joint hybrid and passive beamforming method strikes an attractive trade-off between the bandwidth efficiency and power consumption.
    Our simulation results demonstrate the superiority of the proposed method compared to its traditional counterparts.
	\end{abstract}
	
	\IEEEpeerreviewmaketitle

	\section{Introduction}\label{S1}
    \IEEEPARstart{N}{ext}-generation wireless communication systems tend to aim for Gigabit-per-second data transmission rates \cite{8207426,8922634,9136592} and millimeter-wave (mmWave) techniques are indeed capable of supporting such high data rates given their abundant bandwidth \cite{6979963,8241348,6824746,Access_TSRappaport_mmwave}.
    For compensating the high path-loss of mmWave signals, multi-input multi-output (MIMO) techniques are widely used \cite{7400949,8337813,9007529,7306538}.
    For instance, beamforming is capable of achieving a high directional gain.
    However, due to the high directivity, beamformed mmWave signals can be easily blocked by obstacles \cite{wang2020joint}.
    As a remedy, reconfigurable intelligent surfaces (RISs) can be employed for reflecting the incident signals by intelligently tuning the passive elements embedded into their surfaces \cite{wang2020joint,9475159,9305278,8811733,8930608,9226616,9226135,9409636,9525086,9371019}.
    {Note that full-duplex relaying is also capable of compensating the significant attenuation of mmWave signals \cite{8340227,9337215,9263315}.
    It can also be employed for improving the system's converge without reducing the transmission rate\cite{9115248}.
    However, compared to full-duplex relays, RIS does not need an active transmitter module and only reflects the received signals in a passive way, which incurs no transmit power consumption \cite{8811733}.
    Moreover, full-duplex relays require sophisticated self-interference cancellation, while a RIS naturally operates in full-duplex mode without self-interference or without introducing extra thermal noise \cite{9475160,9110863,9490324}.}
    Moreover, since the mmWave channel exhibits a sparse channel impulse response (CIR), this may be readily exploited for reducing the complexity by hybrid beamforming structures at the base station (BS) \cite{el2013spatially,7579557,7160780}.
    Explicitly, upon appropriately designing the hybrid beamforming at the BS and the passive beamforming at the RIS, RIS-assisted mmWave communication systems exhibit excellent performance at low hardware cost and low power consumption.

    \subsection{Literature overview}\label{S1.1}
    There is now quite a bit of literature on the joint active and passive beamforming design conceived for maximizing the system's bandwidth efficiency, where fully-digital structures are considered at the BS.
    The authors of \cite{8982186,9320618,9117136,li2020intelligent,9076830} propose to jointly design the active and passive beamforming matrices of both RIS-assisted single-input single-output (SISO) and multi-user multi-input single-output (MU-MISO) systems.
    Specifically, Guo \textit{et al.} \cite{8982186} propose a fractional programming (FP)-based framework for joint active and passive beamforming design, while Ma \textit{et al.} \cite{9320618} use FP for solving the bandwidth-efficiency-maximization problem.
    As a further advance, Yan \textit{et al.} \cite{9117136} advocate a sample average approximation (SAA)-based iterative algorithm for their passive beamforming design.
    Li \textit{et al.} \cite{li2020intelligent} exploit the characteristics of RIS elements in a wideband scenario and propose a Lagrangian multiplier based method for jointly designing the active and passive beamforming matrices.
    By contrast, Zhou \textit{et al.} \cite{9076830} conceive a pair of algorithms under the majorization-minimization (MM) algorithmic framework for maximizing the bandwidth efficiency of RIS-assisted multi-group multi-cast MISO systems.
    Furthermore, the authors of \cite{9043523,9110912,9110849,9090356} focus their attention on RIS-assisted MIMO systems, where both the BS and the user are equipped with multiple antennas.
    Specifically, Ning \textit{et al.} \cite{9043523} propose a sum-path-gain maximization (SPGM)-based method for designing the passive beamforming matrix.
    As another valuable contribution, Zhang \textit{et al.} \cite{9110912} derive the closed-form solution of RIS beamforming and propose an alternating optimization (AO)-based method for jointly designing the fully-digital beamforming matrix and the RIS elements.
    To elaborate further, Pan \textit{et al.} \cite{9110849} propose a joint active and passive beamforming design for IRS-assisted simultaneous wireless information and power transfer (SWIPT) systems, where they design the active beamforming by the classic Lagrangian multiplier method, while the passive beamforming is optimized by MM-based and complex circle manifold (CCM)-based methods.
    In a further treatise, Pan \textit{et al.} \cite{9090356} propose to jointly design the fully digital beamforming matrices at the BSs and the RIS elements for their RIS-assisted multi-cell systems, while Zhang \textit{et al.} \cite{2002.03744v2} exploit the FP for jointly designing the active and passive beamforming matrices of RIS-assisted cell-free MIMO systems.

\begin{table*}[t]
    \centering
  \caption{Contrasting our contribution to the literatures}
  \label{tab:Summary}
        \begin{tabular}{| c | c | c | c | c | c | c | c | c |}
        \hline
               & \cite{wang2020joint} & \cite{8982186} & \cite{9110912} & \cite{9219133} & \cite{JSAC_XGao_EnergyEfficient} & \cite{7880698}  & \textbf{Proposed}\\
        \hline
        \hline
            Hybrid Beamforming                   & \checkmark    &            &            & \checkmark & \checkmark & \checkmark & \checkmark \\
        \hline
            RIS                                & \checkmark    & \checkmark & \checkmark &            &            &            & \checkmark \\
        \hline
            Fully-connected hybrid structure   & \checkmark    &            &            & \checkmark &            &            &            \\
        \hline
            Sub-connected hybrid structure     &               &            &            &            & \checkmark & \checkmark & \checkmark \\
        \hline
            Single-resolution phase shifters   & \checkmark    & \checkmark & \checkmark &            & \checkmark & \checkmark &   \\
        \hline
            Twin-resolution phase shifters     &               &            &            & \checkmark &            &            & \checkmark \\
        \hline
        \end{tabular}
\end{table*}

    Moreover, considering the hybrid beamforming structures at the BS, Ning \textit{et al.} \cite{9325920} propose a codebook-based beam training and hybrid beamforming design method for RIS-assisted multi-user MIMO systems.
    In their recent contribution, Wang \textit{et al.} \cite{wang2020joint} minimize the Euclidean distance between the hybrid beamforming matrix and the optimal fully-digital solution derived by singular value decomposition (SVD), and they propose a manifold optimization (MO)-based passive beamforming method for RISs.
    Typically fully-connected hybrid beamforming structures are considered in most of the literatures.
    However, the hardware complexity of the fully-connected hybrid beamforming structure is relatively high.

    {The optimization of different metrics characterizing RIS-assisted systems were also considered.
    The total transmit power minimization problem was investigated in \cite{8811733,8930608}, while the energy efficiency maximization problems of different RIS-assisted systems were considered in \cite{8741198,9352948}.
    Some authors considered the joint optimization of different performance metrics, or striking the best trade-off between bandwidth efficiency and energy efficiency \cite{9309152}.
    Robust beamforming designs were considered in \cite{9180053,9314027,9110587,9133130,9352530,9293148,9449661} in terms of different optimization targets, while the concept of simultaneously transmitting and reflecting RIS (STAR-RIS) was proposed in \cite{9570143}.
    As a further evolution, the active RIS philosophy was proposed in \cite{9377648}, while a hybrid RIS structure was conceived in \cite{nguyen2021hybrid}.
    }

    For hybrid beamforming, one of the important research objectives is to reduce both the hardware complexity and the power consumption of the analog beamformer by using low-resolution phase shifters.
    In this context, Sohrabi and Yu \cite{7389996} propose an iterative hybrid beamforming algorithm for fully-connected hybrid structures having low-resolution phase shifters.
    They further extend the proposed method to wideband scenarios in \cite{7913599}.
    As another state-of-the-art (SoA) contribution, Wang \textit{et al.} \cite{8331836} jointly design the beamformer and combiner relying on low-resolution phase shifters.
    Chen \textit{et al.} \cite{8303761} propose a hybrid beamforming matrix in conjunction with low-resolution phase shifters, where an iterative training-based method is proposed, which converges to the dominant steering vectors that are aligned with the direction of the highest channel gain.
    As a further development, Li \textit{et al.} \cite{9110865} propose a Lagrangian multiplier combined with the penalty dual decomposition (PDD) method for designing a hybrid beamforming matrix realized by low-resolution phase shifters in wideband mmWave systems.
    By relying on both high- and low-resolution phase shifters, in our recent work \cite{9219133} we propose a fully-connected hybrid structure relying on twin-resolution phase shifters,  where we also conceive a dynamic hybrid beamforming method.
    However, the fully-connected analog beamformer suffers from high hardware complexity.
    For mitigating this problem, Gao \textit{et al.} \cite{JSAC_XGao_EnergyEfficient} conceive a pioneering sub-connected analog beamformer conceived with a successive interference cancelation (SIC) based hybrid beamforming method.
    As a beneficial evolution, Park \textit{et al.} \cite{7880698} propose a dynamic sub-connected analog beamformer, where the connection between the radio-frequency (RF) chains and all the phase shifters can be flexibly set by their greedy method.
    In Table \ref{tab:Summary} we boldly and explicitly contrast our contributions to the SoA, emphasizing that our goal is to develop RIS-assisted mmWave systems relying on a low-complexity and energy-efficient hybrid structure, harvesting a joint hybrid and passive beamforming design method.

    \subsection{Our contribution}\label{S1.2}
    Against this background, we propose a RIS-assisted point-to-point mmWave MIMO system relying on twin-resolution dynamic sub-connected hybrid beamforming structures.
    The detailed contributions of this paper are summarized as follows:
\begin{enumerate}
    \item
    An energy-efficient and low-complexity dynamic sub-connected analog beamformer realized by twin-resolution phase shifters is proposed, where half of the phase shifters in each sub-array are predefined to have high resolution, while others have low resolution.
    The connection between the twin-resolution phase shifters and the TAs in each sub-array can be flexibly determined according to the channel state information (CSI).
    The proposed analog beamformer has lower hardware complexity than the fully-connected analog beamformer of \cite{9219133}.
    Moreover, it can strike a more attractive bandwidth efficiency vs. power consumption tradeoff than the conventional sub-connected analog beamformer of \cite{JSAC_XGao_EnergyEfficient}.
    \item
    We propose to maximize the bandwidth efficiency by jointly optimizing the hybrid beamforming at the BS and passive beamforming realized by the RIS elements.
    However, both the objective function (OF) and constant-modulus constraints imposed on each element of the analog beamforming matrix and on the RIS elements are non-convex, which makes the problem formulated challenging to solve.
    Inspired by the successive interference cancelation (SIC)-based algorithm, we first transform the bandwidth efficiency maximization problem into a series of sub-rate maximization problems, where the sub-array of phase shifters and RIS elements are jointly optimized.
    Then we propose a SIC-based joint hybrid and passive beamforming algorithm to solve them.
    \item
    For the hybrid beamforming design, the well-known SVD-based digital beamformer component is used.
    Then we design the analog beamformer component column-by-column by relying on SIC.
    During the design of each sub-array, we fix the RIS elements and regard the cascaded BS-RIS-user link as the effective channel.
    We propose a greedy method for solving the corresponding sub-problem.
    \item
    Having the designed sub-arrays, we propose to update the RIS elements based on CCM.
    Specifically, we maximize the upper bound of the bandwidth efficiency of the sub-arrays and RIS elements that have been obtained.
    Then, complex matrix transformations are adopted for transforming the reformulated problem into a tractable form.
    Then we exploit a CCM-based method for updating the passive beamforming matrix.
    Our simulation results demonstrate the superiority of the proposed method over its counterparts.
    \end{enumerate}

    The remainder of this paper is organized as follows.
    In Section \ref{S2}, our downlink system model and channel model are introduced.
    In Section \ref{S3}, our twin-resolution dynamic sub-connected analog beamformer and problem formulation are presented.
    In Section \ref{S4}, our proposed joint hybrid and passive beamforming design is introduced.
    In Section \ref{S5}, our numerical results are provided.
    Finally, our conclusions are drawn in Section \ref{S6}.

    \emph{Notation}:
	Lower-case and upper-case boldface letters denote vectors and matrices, respectively;
	$(\cdot)^{*}$,$(\cdot)^{\rm{T}}$, $(\cdot)^{\text{H}}$, $(\cdot)^{-1}$ and $(\cdot)^{\dagger}$ denote the conjugate, transpose, conjugate transpose, inverse and pseudo-inverse of a matrix, respectively;
    $\mathrm{Tr}(\cdot)$ presents the trace function;
    $\mathrm{diag}\left[a_1,a_2,\cdots,a_N\right]$ denotes a diagonal matrix with $a_1,a_2,\cdots,a_N$ being the diagonal elements;
    $\| \cdot \|_{\rm F}$ denotes the Frobenius norm of a matrix;
    $|a|$ is the absolute value of a scalar;
    $|\mathbf{A}|$ is the determinant of a matrix;
    $\mathbf{A}{\left(i,:\right)}$ and $\mathbf{A}{\left(:,j\right)}$ represent the $i$-th row and $j$-th column of the matrix $\mathbf{A}$, respectively;
    The operator $\circ$ represents the Hadamard product;
	Finally, $\mathbf{I}_P$ denotes the identity matrix of size $P\times P$.
	\section{Downlink System Model and Channel Model}\label{S2}
 	\begin{figure}[t]
		\center{\includegraphics[width=0.5\textwidth]{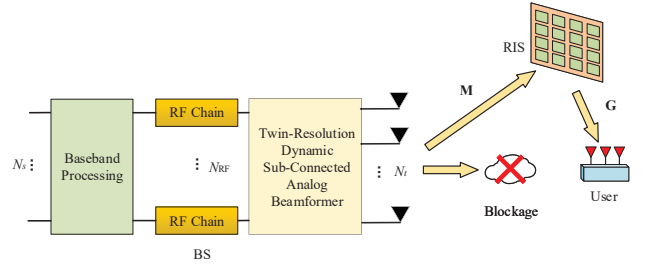}}
		\caption{Illustration of the RIS-assisted mmWave MIMO systems relying on twin-resolution dynamic sub-connected hybrid structure.}
		\label{system_model}
	\end{figure}
    In this section, we will introduce both the system model and channel model of our RIS-assisted point-to-point mmWave MIMO downlink system relying on twin-resolution dynamically reconfigurable sub-connected hybrid structures.
	\subsection{RIS-assisted Point-to-point MmWave MIMO Downlink Model}\label{S2.1}
    As shown in Fig. \ref{system_model}, the BS communicates with the user through the reflected link (BS-RIS-user), since the direct link (BS-user) is assumed to be blocked.
    The BS is equipped with hybrid beamforming relying on $N_t$ transmit antennas (TAs) and $N_{\rm RF}$ RF chains.
    At the user, the fully-digital structure having $N_r$ receive antennas (RAs) is adopted.
    The transmit signal at the BS has $N_s$ data streams, which is defined as $\mathbf{\mathbf{s}} = \left[ s_1,s_2,\cdots,s_{N_s} \right]^{\text{T}} \in \mathbb{C}^{N_s \times 1}$ and has the normalized power of $\mathbb{E}\left[{\mathbf{s}\mathbf{s}^{\text{H}}}\right] = \mathbf{I}_{N_s}$.
    At the BS, the transmit signals are firstly precoded by the baseband digital beamformer $\mathbf{F}_{\mathrm{BB}} \in \mathbb{C}^{N_{\mathrm{RF}} \times N_s}$.
    Without loss of generality, we assume $N_s = N_{\rm RF}$ \cite{JSAC_XGao_EnergyEfficient,han2015large}.
    Then the signals $\mathbf{F}_{\mathrm{BB}} \mathbf{s}$ are precoded by the analog beamformer $\mathbf{F}_{\mathrm{RF}} \in \mathbb{C}^{ N_t \times N_{\mathrm{RF}}}$, which will be introduced in detail in Section \ref{S3.1}.
    The phase shifters in the analog beamformer have the same amplitude of $\frac{1}{\sqrt{M}}$ with $M = \frac{N_t}{N_{\rm RF}}$ representing the number of phase shifters in each sub-array.
    Hence, we have
    \begin{align}\label{F_RF}
    \mathbf{F}_{\mathrm{RF}} =
    \begin{bmatrix} \mathbf{f}_{\mathrm{RF},1} & \mathbf{0} & \cdots & \mathbf{0} \\
    \mathbf{0} & \mathbf{f}_{\mathrm{RF},2} & \cdots & \mathbf{0} \\
    \mathbf{0} & \mathbf{0} & \cdots & \mathbf{f}_{\mathrm{RF},N_{\rm RF}}
    \end{bmatrix},
	\end{align}
    where $\mathbf{f}_{\mathrm{RF},j} \in \mathbb{C}^{M \times 1}$ and $\mathbf{f}_{\mathrm{RF},j}\left(i\right) = \frac{1}{\sqrt{M}}e^{j\theta_{i,j}}$ with $\theta_{i,j}$ denoting the discrete phase shift caused by the limited-resolution phase shifters.
    The total transmit power constraint is set to $\left\| \mathbf{F}_{\mathrm{RF}}\mathbf{F}_{\mathrm{BB}} \right\|_{\rm F}^2 = N_{\rm RF} $.
    The transmitted signal after hybrid analog and digital beamforming is expressed by
    \begin{align}\label{s_BS}
    \mathbf{x} = \sqrt{\frac{P_t}{N_s}}\mathbf{F}_{\mathrm{RF}}\mathbf{F}_{\mathrm{BB}}\mathbf{s},
	\end{align}
    where $P_t$ is the transmit power.

    Then, the transmitted signals pass through the cascaded BS-RIS-user channel, which is defined as $\mathbf{H}_{\rm eff} = \mathbf{G}\mathbf{\Phi}\mathbf{M}$, where $\mathbf{M} \in \mathbb{C}^{N_{\rm RIS} \times N_t}$ is the BS-RIS channel, $\mathbf{G} \in \mathbb{C}^{N_r \times N_{\rm RIS}}$ is the RIS-user channel, and $\mathbf{\Phi} = \text{diag}\left[ e^{j\phi_1},e^{j\phi_2},\cdots,e^{j\phi_{N_{\rm RIS}}} \right] \in \mathbb{C}^{N_{\rm RIS} \times N_{\rm RIS}}$ is the RIS matrix having $N_{\rm RIS}$ passive elements.
    {It is assumed that the channel state information (CSI) is perfectly known at the BS, noting that both the accurate channel estimation and the robust joint hybrid and passive design relying on partial CSI constitute rather specific problems in RIS-assisted systems relying on dynamically configured subarrays \cite{9110869,alexandropoulos2017robustness,zhou2021channel,9614196,9241029}.}
    Then the signal $\mathbf{y} \in \mathbb{C}^{N_r \times 1}$ received by the user can be expressed as
    \begin{align}\label{y}
    \mathbf{y} = \sqrt{\frac{P_t}{N_s}}\mathbf{G}\mathbf{\Phi}\mathbf{M}\mathbf{F}_{\mathrm{RF}}\mathbf{F}_{\mathrm{BB}}\mathbf{s} + \mathbf{n},
	\end{align}
    where $ \mathbf{n} \sim \mathcal{CN}\left(0, \sigma^2\mathbf{I}_{N_r} \right) \in \mathbb{C}^{N_{r} \times 1}$ denotes the additive white Gaussian noise.
    We can further express the achievable bandwidth efficiency as \cite{wang2020joint}
    \begin{align}\label{R}
    R = \mathrm{log}_2 & \left(\left| \mathbf{I}_{N_r} + \frac{P_t}{\sigma^2 N_s}  \mathbf{G}\mathbf{\Phi}\mathbf{M} \mathbf{F}_{\mathrm{RF}} \mathbf{F}_{\mathrm{BB}}\cdot \right.\right.\nonumber\\
    & \quad \quad \quad \quad \quad \quad \left.\left. \mathbf{F}^{\text{H}}_{\mathrm{BB}} \mathbf{F}^{\text{H}}_{\mathrm{RF}} \mathbf{M}^{\text{H}} \mathbf{\Phi}^{\text{H}} \mathbf{G}^{\text{H}} \right|\right).
	\end{align}
    \subsection{Channel Model}\label{S2.2}
    In this paper, we adopt the multi-path mmWave channel model for both the BS-RIS channel $\mathbf{M}$ and the RIS-user channel $\mathbf{G}$, which are defined by \cite{el2013spatially}
    \begin{align}\label{M}
    \mathbf{M} = \sqrt{\frac{N_t N_\mathrm{RIS}}{L}} \sqrt{A_{\rm BR}} \sum_{\ell = 1}^{L}\alpha_{\ell}\mathbf{a}_{\mathrm{RIS}}\left( \phi^{\mathrm{r}}_{\ell}, \theta^{\mathrm{r}}_{\ell} \right)\mathbf{a}_{\mathrm{BS}}^{\text{H}}\left( \phi^{\mathrm{t}}_{\ell}, \theta^{\mathrm{t}}_{\ell} \right),
	\end{align}
    \begin{align}\label{G}
    \mathbf{G} = \sqrt{\frac{N_\mathrm{RIS}N_r}{P}} \sqrt{A_{\rm RU}} \sum_{p = 1}^{P}\beta_{p}\mathbf{a}_{\mathrm{MS}}\left( \phi^{\mathrm{r}}_{p}, \theta^{\mathrm{r}}_{p} \right)\mathbf{a}_{\mathrm{RIS}}^{\text{H}}\left( \phi^{\mathrm{t}}_{p}, \theta^{\mathrm{t}}_{p} \right),
	\end{align}
    where $L$ and $P$ denote the total number of paths in the BS-RIS link and RIS-user link, respectively.
    Symbols $A_{\rm BR}$ and $A_{\rm RU}$ represent the path-loss in the BS-RIS link and RIS-user link, respectively.
    The variables $\alpha_l, \beta_p \sim \mathcal{CN}\left(0, 1 \right)$ denote the complex gain in the BS-RIS link and RIS-user link, respectively.
    The variables $\phi^{\mathrm{t}}_{\ell}$ $\left(\phi^{\mathrm{r}}_{\ell}\right)$ and $\theta^{\mathrm{t}}_{\ell}$ $\left(\theta^{\mathrm{r}}_{\ell}\right)$ are the azimuth and elevation angles of departure (arrival) associated with the BS-RIS link, $\phi^{\mathrm{t}}_{p}$ $\left(\phi^{\mathrm{r}}_{p}\right)$ and $\theta^{\mathrm{t}}_{p}$ $\left(\theta^{\mathrm{r}}_{p}\right)$ represent the azimuth and elevation angles of departure (arrival) associated with the RIS-user link.
    Vector $\mathbf{a}_{o}$, $o \in \left\{ \mathrm{BS},\mathrm{RIS},\mathrm{MS} \right\}$ stands for the beam steering vectors at the BS, the RIS or the user.
    Uniform planar arrays (UPAs) of antennas are employed at the BS, the RIS and the user.
    Thus, the typical beam steering vector of UPA antennas can be expressed by
    \begin{align}\label{a_BS}
    \mathbf{a}_o\left( \phi, \theta \right) = & \frac{1}{\sqrt{WH}}  \left[ 1,\cdots, e^{j\frac{2\pi}{\lambda}d\left( m \mathrm{sin} \phi \mathrm{sin} \theta + n \mathrm{cos}\theta \right) } ,\cdots, \right. \nonumber\\
    & \left. e^{j\frac{2\pi}{\lambda}d\left( \left( W-1 \right) \mathrm{sin} \phi \mathrm{sin} \theta + \left( H-1 \right) \mathrm{cos}\theta \right) } \right]^{\text{T}},
	\end{align}
    where $m \in \{0,1,\cdots,W-1 \}$ and $n \in \{0,1,\cdots,H-1 \}$ with $W$ and $H$ denoting the number of antennas in the horizontal and vertical directions.
%    \subsection{Problem Formulation}\label{S2.3}
    \section{Proposed Twin-Resolution Dynamic Sub-Connected Analog Beamformer and Problem Formulation}\label{S3}
    In this section, we will firstly introduce the structure of our twin-resolution dynamically reconfigurable sub-connected analog beamformer, followed by the associated optimization problem formulation.
    \subsection{Proposed Twin-Resolution Dynamic Sub-Connected Analog Beamformer}\label{S3.1}
 	\begin{figure}[t]
		\center{\includegraphics[width=0.4\textwidth]{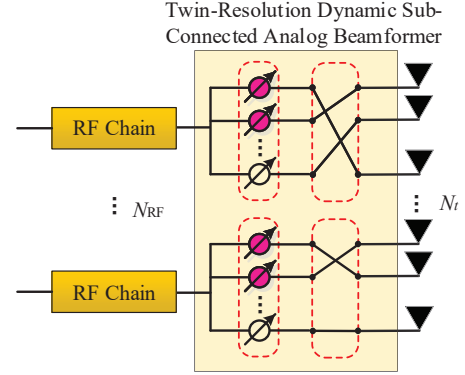}}
		\caption{Illustration of the proposed twin-resolution dynamic sub-connected analog beamformer.}
		\label{Analog}
	\end{figure}
    As illustrated in Fig. \ref{Analog}, we propose a twin-resolution dynamic sub-connected analog beamformer realized by twin-resolution phase shifters.
    Specifically, we predefine half of the phase shifters in each sub-array as high-resolution ones (depicted as solid circles in Fig. \ref{Analog}), while the others as low-resolution ones (depicted as hollow circles in Fig. \ref{Analog}).
    Therefore, the number of high-resolution and low-resolution phase shifters in each sub-array is $N_{{\rm H}} = N_{{\rm L}} = \frac{M}{2}$.
%    For intuitive expression, we connect $M$ nodes to each RF chain.
    For achieving near-optimal performance, the connection between the $M$ phase shifters having different resolutions and $M$ TAs can be flexibly designed through switches according to the CSI.
    The phase shifter connected to the $m$-th TA in the $j$-th sub-array corresponds to the $m$-th element in $\mathbf{f}_{\mathrm{RF},j}$.
    By beneficially selecting the resolution of the phase shifter connected to each TA and designing the phase of the selected high-resolution or low-resolution phase shifter, we arrive at an attractive analog beamformer design.
    This procedure will be detailed in Section \ref{S4.1}.
%     	\begin{figure}[t]
%		\center{\includegraphics[width=0.4\textwidth]{switch_per_phase_shift.eps}}
%		\caption{The switch network between one phase shifter and the TAs in one sub-array.}
%		\label{connection_one}
%	\end{figure}

    The novelty of the proposed twin-resolution dynamic sub-connected analog beamformer lies in two aspects.
    Firstly, compared to our recently proposed twin-resolution fully-connected analog beamformer \cite{9219133}, the twin-resolution dynamic sub-connected analog beamformer has lower hardware complexity.
    Specifically, the total number of switches required for the hybrid beamformer is $\left({\frac{N_t}{N_{\rm RF}}}\right)^2N_{\rm RF} = \frac{N_t^2}{N_{\rm RF}}$ in the proposed analog beamformer component.
    This is significantly lower than that of a twin-resolution fully-connected analog beamformer, which is $N_t^2N_{\rm RF}$, especially when the number of RF chains is large.
    Secondly, compared to the classical sub-connected analog beamformer of \cite{JSAC_XGao_EnergyEfficient}, our proposed scheme is more energy efficient and can strike an attractive bandwidth efficiency vs. energy efficiency trade-off, since the high-resolution phase shifters can attain a high beamforming gain, while the low-resolution phase shifters deployed in the analog beamformer dissipate much less energy.

    \subsection{Problem Formulation}\label{S3.2}
    In this paper, we propose to jointly optimize the hybrid beamforming matrix at the BS and the passive beamforming at the RIS for achieving the maximum bandwidth efficiency.
    The achievable bandwidth-efficiency maximization problem is formulated as
    \begin{subequations}\label{ObjSE_v1}
    \begin{align}
    \max_{ \mathbf{F}_{\rm RF},\mathbf{F}_{\rm BB},\mathbf{\Phi} } & \quad  \quad\quad   \quad R \\
    s.t. \quad  \quad & \left\| \mathbf{F}_{\mathrm{RF}}\mathbf{F}_{\mathrm{BB}} \right\|_{\rm F}^2 = N_{\rm RF} \\
    \quad  \quad & \left|\mathbf{f}_{\mathrm{RF},j}\left(i\right)\right| = \frac{1}{\sqrt{M}}, \nonumber\\
    & \quad  \quad  \forall i = 1,2,\cdots,M, \forall j = 1,2,\cdots,N_{\rm RF}\\
    \quad  \quad & \theta_{i,j} \in \mathcal{Q}_{\rm H} \text{\ or } \mathcal{Q}_{\rm L}, \nonumber \\
    & \quad  \quad \forall i = 1,2,\cdots,M, \forall j = 1,2,\cdots,N_{\rm RF}\\
    \quad  \quad & N_{\rm H} = N_{\rm L} = \frac{M}{2}\\
    \quad  \quad & \mathbf{\Phi} = \text{diag}\left[ e^{j\phi_1},e^{j\phi_2},\cdots,e^{j\phi_{N_{\rm RIS}}} \right].
    \end{align}
    \end{subequations}
    We observe that problem (\ref{ObjSE_v1}) is a non-convex optimization problem due to the constant-modulus constraints (\ref{ObjSE_v1}c) and (\ref{ObjSE_v1}f).
    Moreover, the hybrid beamforming matrix $\mathbf{F}_{\rm RF}\mathbf{F}_{\rm BB}$ and passive beamforming matrix $\mathbf{\Phi}$ are highly coupled with each other, which makes this problem quite challenging to solve.
    Hence we will propose a SIC-based joint design method for solving this problem.
    \subsection{Sub-Problem Formulation}\label{S4.1}
    Inspired by the SIC scheme of \cite{JSAC_XGao_EnergyEfficient}, we first derive the sub-problem for each sub-array.
    Specifically, the objective function (OF) of (\ref{ObjSE_v1}a) is rewritten as
    \begin{align}\label{ObjSE_v2}
    R = \mathrm{log}_2\left(\left| \mathbf{I}_{N_r} + \frac{P_t}{\sigma^2 N_s}  \mathbf{H}_{\rm eff} \mathbf{F}_{\rm RF}\mathbf{F}_{\rm BB} \mathbf{F}_{\rm BB}^{\text{H}}  \mathbf{F}_{\rm RF}^{\text{H}} \mathbf{H}_{\rm eff}^{\text{H}} \right|\right),
    \end{align}
    where $\mathbf{H}_{\rm eff} = \mathbf{G}\mathbf{\Phi}\mathbf{M}$ denotes the effective channel matrix.
    Then we calculate the digital precoding matrix $\mathbf{F}_{\mathrm{BB}} = \delta \left( \mathbf{F}_{\mathrm{RF}}^{\rm H}\mathbf{F}_{\mathrm{RF}} \right)^{-\frac{1}{2}}\mathbf{V}_{\rm eff}$, where $\mathbf{V}_{\rm eff}$ represents the first $N_{\rm RF}$ right singular vectors of $\mathbf{H}_{\rm eff} \mathbf{F}_{\rm RF}$, and $\delta = \sqrt{\frac{N_s}{\left\| \mathbf{F}_{\mathrm{RF}}\mathbf{F}_{\mathrm{BB}} \right\|_{\rm F}^2}}$ is a coefficient used for satisfying the total transmit power constraint \cite{8902153}.
    Substituting $\mathbf{F}_{\mathrm{BB}}$ into $\delta$, we have $\delta = 1$.
    Then, we rewrite the analog beamforming matrix as
    \begin{align}\label{F}
    \mathbf{F}_{\rm RF} & =
    \begin{bmatrix} \mathbf{F}_{\mathrm{RF},1} & \mathbf{F}_{\mathrm{RF},2} & \cdots & \mathbf{F}_{\mathrm{RF},N_{\rm RF}}
    \end{bmatrix}\nonumber\\
    & =
    \begin{bmatrix} \mathbf{f}_{\mathrm{RF},1} & \mathbf{0} & \cdots & \mathbf{0} \\
    \mathbf{0} & \mathbf{f}_{\mathrm{RF},2} & \cdots & \mathbf{0} \\
    \mathbf{0} & \mathbf{0} & \cdots & \mathbf{f}_{\mathrm{RF},N_{\rm RF}}
    \end{bmatrix}.
	\end{align}
    After some mathematical transformations, we can equivalently rewrite (\ref{ObjSE_v2}) as \cite{JSAC_XGao_EnergyEfficient}
    \begin{align}\label{R_SIC}
    R = \sum_{j=1}^{N_{\rm RF}} \mathrm{log}_2 & \left(\left| 1 + \frac{P_t }{\sigma^2 N_s} \mathbf{F}_{\mathrm{RF},j}^{\rm H}\mathbf{H}_{\rm eff}^{\text{H}} \mathbf{T}_{j}^{-1} \mathbf{H}_{\rm eff} \mathbf{F}_{\mathrm{RF},j} \right|\right),
	\end{align}
    where $\mathbf{T}_{1} = \mathbf{I}_{N_r}$, $\mathbf{T}_{j} = \mathbf{I}_{N_r} + \frac{P_t}{\sigma^2 N_s} \mathbf{H}_{\rm eff} \overline{\mathbf{F}}_{\mathrm{RF},j-1}  \overline{\mathbf{F}}_{\mathrm{RF},j-1}^{\text{H}} \mathbf{H}_{\rm eff}^{\text{H}}$ for $ j \geq 2$, and $\overline{\mathbf{F}}_{\mathrm{RF},j-1} \in \mathbb{C}^{N_t \times \left(j-1\right)}$ is composed of the first $j-1$ columns of $\mathbf{F}_{\mathrm{RF}} $ \cite{JSAC_XGao_EnergyEfficient}.

    %In this paper, we propose to update the passive beamforming matrix $\mathbf{\Phi}$ after the design of each sub-array.
%    Thus, we introduce the passive beamforming matrix $\mathbf{\Phi}_{j-1}$ for the joint hybrid and passive beamforming design of the $j$-th sub-array.
    {In this paper, we propose to update the passive beamforming matrix $\mathbf{\Phi}$ after the design of each sub-array.
    Thus, we define the matrix $\mathbf{\Phi}$ obtained by the CCM-based method after the design of the $j$-th sub-array as $\mathbf{\Phi}_j$.
    We can arrive at the final $\mathbf{\Phi} = \mathbf{\Phi}_{N_{\rm RF}}$ after the design of the last sub-array, which is shared by all RF chains.
    Hence the effective channel in each sub-problem is $ \mathbf{H}_{{\rm eff},j} = \mathbf{G}\mathbf{\Phi}_{j-1}\mathbf{M} $ with $\mathbf{\Phi}_{j-1}$ denoting the passive beamforming matrix obtained according to the design of the previous sub-arrays.
    %Hence the equivalent channel is correspondingly reformulated as $\mathbf{H}_{{\rm eff},j} = \mathbf{G}\mathbf{\Phi}_{j-1}\mathbf{M}$.
    Furthermore, we can divide problem (\ref{ObjSE_v1}) into $N_{\rm RF}$ sub-problems, where the $j$-th sub-problem $\left( j = 1,2,\cdots,N_{\rm RF} \right)$ is formulated as }
    \begin{subequations}\label{ObjSE_v3}
    \begin{align}
    \max_{ \mathbf{f}_{\mathrm{RF},j},\mathbf{\Phi}_j } \quad & \mathrm{log}_2\left(\left| 1 + \frac{P_t }{\sigma^2 N_s} \mathbf{f}_{\mathrm{RF},j}^{\rm H} \mathbf{B}_{j} \mathbf{f}_{\mathrm{RF},j} \right|\right) \\
    s.t.  \quad  \quad & \left|\mathbf{f}_{\mathrm{RF},j}\left(i\right)\right| = \frac{1}{\sqrt{M}}, \forall i = 1,2,\cdots,M\\
    \quad  \quad & \theta_{i,j} \in \mathcal{Q}_{\rm H} \text{\ or } \mathcal{Q}_{\rm L}, \forall i = 1,2,\cdots,M\\
    \quad  \quad & N_{\rm H} = N_{\rm L} = \frac{M}{2}\\
    \quad  \quad & \mathbf{\Phi}_j = \text{diag}\left[ e^{j\phi_1},e^{j\phi_2},\cdots,e^{j\phi_{N_{\rm RIS}}} \right],
    \end{align}
    \end{subequations}
    where $\mathbf{B}_{j} \in \mathbb{C}^{M \times M}$ is composed of the rows and columns of $\mathbf{H}_{{\rm eff},j}^{\text{H}} \overline{\mathbf{T}}_{j}^{-1} \mathbf{H}_{{\rm eff},j}$ from the $\left(M\left( j-1 \right)\right. $ $\left.+1 \right)$-st one to the $\left(Mj\right)$-th one, $\overline{\mathbf{T}}_{1} = \mathbf{I}_{N_r}$, $\overline{\mathbf{T}}_{j} = \mathbf{I}_{N_r} + \frac{P_t}{\sigma^2 N_s} \mathbf{H}_{{\rm eff},j} \overline{\mathbf{F}}_{\mathrm{RF},j-1}  \overline{\mathbf{F}}_{\mathrm{RF},j-1}^{\text{H}} \mathbf{H}_{{\rm eff},j}^{\text{H}}$ for $ j \geq 2$.

{According to Problem (\ref{ObjSE_v3}), we observe that the bandwidth efficiency maximization problem can be transformed into a series of sub-rate maximization problems, each of which is for a specific sub-array.
Then we optimize the sub-problems one by one \cite{JSAC_XGao_EnergyEfficient}.
Note that the specific order of the design has no influence on the bandwidth efficiency, which is determined by the sum of the sub-rates of all sub-arrays.
By relying on an SIC scheme, we maximize the sub-rate of the first sub-array by jointly designing the twin-resolution phase shifters and the RIS elements.
Then we obtain $\mathbf{f}_{\mathrm{RF},1}$ and $\mathbf{\Phi}_1$ for updating the matrix $\mathbf{B}_{2}$, followed by similar procedures for the subsequent design steps, optimizing the sub-rate of the remaining sub-arrays one by one, until the last sub-array is considered.
}
    \section{Proposed Joint Hybrid and Passive Beamforming Design}\label{S4}
    In this section, we will propose solutions to problem (\ref{ObjSE_v3}) obtained by a SIC-based joint design method.

    \subsection{Greedy Method Proposed for Hybrid Beamforming Design}\label{S4.2}
    Relying on SIC, we design the analog beamformer $\mathbf{F}_{\rm RF}$ column-by-column.
    For designing the $j$-th column $\mathbf{F}_{{\rm RF},j}$ of $\mathbf{F}_{\rm RF}$, we assume a fixed passive beamforming matrix $\mathbf{\Phi}_{j-1}$.
    Problem (\ref{ObjSE_v3}) can be reformulated as
    \begin{subequations}\label{ObjSE_v3_HB}
    \begin{align}
    \max_{ \mathbf{f}_{\mathrm{RF},j} } \quad & \mathrm{log}_2\left(\left| 1 + \frac{P_t}{\sigma^2 N_s} \mathbf{f}_{\mathrm{RF},j}^{\rm H} \mathbf{B}_{j} \mathbf{f}_{\mathrm{RF},j} \right|\right) \\
    s.t.  \quad  \quad & \left|\mathbf{f}_{\mathrm{RF},j}\left(i\right)\right| = \frac{1}{\sqrt{M}}, \forall i = 1,2,\cdots,M\\
    \quad  \quad & \theta_{i,j} \in \mathcal{Q}_{\rm H} \text{\ or } \mathcal{Q}_{\rm L}, \forall i = 1,2,\cdots,M\\
    \quad  \quad & N_{\rm H} = N_{\rm L} = \frac{M}{2}.
    \end{align}
    \end{subequations}
{Note that the conventional phase shifter design in the sub-connected structure of \cite{JSAC_XGao_EnergyEfficient} and in the twin-resolution fully-connected structure of \cite{9219133} are not suitable for the proposed twin-resolution dynamically sub-connected solution.
    Therefore, we propose a greedy method for determining the connections between the twin-resolution phase shifters and TAs in each sub-array.}
    The initialization of $\mathbf{f}_{\mathrm{RF},j}$ is set to $\mathbf{f}_{\mathrm{RF},j} = \frac{1}{\sqrt{M}}e^{j\text{angle}\left( \mathbf{v}_j \right)}$ \cite{JSAC_XGao_EnergyEfficient}, where $\mathbf{v}_j$ is the first right singular vector of $\mathbf{B}_{j}$.
    We also initialize two counters $c_{\rm H}=c_{\rm L}=0$ for counting the number of twin-resolution phase shifters that have been connected to TAs.
    We define the set of discrete phases for the high-resolution and low-resolution phase shifters as $\mathcal{Q}_{\mathrm{H}}$ and $\mathcal{Q}_{\mathrm{L}}$, respectively.
    Then we calculate $q_{\rm L}\left( i \right) = \min_{q\in Q_{\rm L}}\left|{\theta_{i,j} - Q_{\rm L}\left\{ q \right\}}\right|$ and $q_{\rm H}\left( i \right) = \min_{q\in Q_{\rm H}}\left|{\theta_{i,j} - Q_{\rm H}\left\{ q \right\}}\right|$ for $\forall i \in \mathcal{S}_j$, where $\mathcal{S}_j = \left\{ 1,2,\cdots,M \right\}$ is defined as the index set of TAs in the $j$-th sub-array.
    This step is used for obtaining the quantization errors for all elements in $\mathbf{f}_{\mathrm{RF},j}$.
    Subsequently, we denote the indices of the element in $\mathbf{f}_{\mathrm{RF},j}$ having the minimal quantization error for the high-resolution and low-resolution phase shifters as $i_{\rm L} = \min_{i\in \mathcal{S}_j}\left|{\theta_{i,j} - Q_{\rm L}\left\{ q_{\rm L}\left( i \right) \right\}}\right|$ and $i_{\rm H} = \min_{i\in \mathcal{S}_j}\left|{\theta_{i,j} - Q_{\rm H}\left\{ q_{\rm H}\left( i \right) \right\}}\right|$ for $\forall i \in \mathcal{S}_j$.
    We check the relationship between $\left|{\theta_{i_{\rm L},j} - Q_{\rm L}\left\{ q_{\rm L}\left( i_{\rm L} \right) \right\}}\right|$ and $\left|{\theta_{i_{\rm H},j} - Q_{\rm H}\left\{ q_{\rm H}\left( i_{\rm H} \right) \right\}}\right|$ and determine the connection between a specific TA and a phase shifter as follows.
    \begin{enumerate}[\text{Case} 1: ]
    \item
    If $\left|{\theta_{i_{\rm L},j} - Q_{\rm L}\left\{ q_{\rm L}\left( i_{\rm L} \right) \right\}}\right| < \left|{\theta_{i_{\rm H},j} - Q_{\rm H}\left\{ q_{\rm H}\left( i_{\rm H} \right) \right\}}\right|$ and $c_{\rm L} < \frac{M}{2}$, we connect the $i_{\rm L}$-th TA in the $j$-th sub-array to a low-resolution phase shifter and set $\mathbf{f}_{\mathrm{RF},j}\left( i_{\rm L} \right) = \frac{1}{\sqrt{M}}e^{jQ_{\rm L}\left\{ q_{\rm L}\left( i_{\rm L} \right) \right\}}$.
    We update $c_{\rm L} = c_{\rm L} + 1$ and $\mathcal{S}_j = \mathcal{S}_j\backslash \left\{ i_{\rm L} \right\}$.
    When $c_{\rm L}=\frac{M}{2}$, we connect the $i_{\rm H}$-th TA in the $j$-th sub-array to a high-resolution phase shifter and set $\mathbf{f}_{\mathrm{RF},j}\left( i_{\rm H} \right) = \frac{1}{\sqrt{M}} e^{jQ_{\rm H}\left\{ q_{\rm H}\left( i_{\rm H} \right) \right\}}$.
    Similarly, we update $c_{\rm H} = c_{\rm H} + 1$ and $\mathcal{S}_j = \mathcal{S}_j\backslash \left\{ i_{\rm H} \right\}$.
    \item
    If $\left|{\theta_{i_{\rm L},j} - Q_{\rm L}\left\{ q_{\rm L}\left( i_{\rm L} \right) \right\}}\right| \geq \left|{\theta_{i_{\rm H},j} - Q_{\rm H}\left\{ q_{\rm H}\left( i_{\rm H} \right) \right\}}\right|$ and $c_{\rm H} < \frac{M}{2}$, the $i_{\rm H}$-th TA in the $j$-th sub-array is connected to a high-resolution phase shifter and set $\mathbf{f}_{\mathrm{RF},j}\left( i_{\rm H} \right) = \frac{1}{\sqrt{M}}e^{jQ_{\rm H}\left\{ q_{\rm H}\left( i_{\rm H} \right) \right\}}$.
    We update $c_{\rm H} = c_{\rm H} + 1$ and $\mathcal{S}_j = \mathcal{S}_j\backslash \left\{ i_{\rm H} \right\}$.
    When $c_{\rm H}=\frac{M}{2}$, the $i_{\rm L}$-th TA in the $j$-th sub-array is connected to a low-resolution phase shifter and $\mathbf{f}_{\mathrm{RF},j}\left( i_{\rm L} \right)$ is set to $\mathbf{f}_{\mathrm{RF},j}\left( i_{\rm L} \right) = \frac{1}{\sqrt{M}}e^{jQ_{\rm L}\left\{ q_{\rm L}\left( i_{\rm L} \right) \right\}}$.
    We update $c_{\rm L} = c_{\rm L} + 1$ and $\mathcal{S}_j = \mathcal{S}_j\backslash \left\{ i_{\rm L} \right\}$.
    \end{enumerate}

    For the subsequent design of the $j$-th sub-array in the twin-resolution dynamic sub-connected analog beamformer, we embark on further derivation of the OF (\ref{ObjSE_v3}a).
    An element-wise decomposition is given by
    \begin{align}\label{decomp_1}
    &\mathrm{log}_2\left(\left| 1 + \frac{P_t }{\sigma^2 N_s} \mathbf{f}_{\mathrm{RF},j}^{\rm H} \mathbf{B}_{j} \mathbf{f}_{\mathrm{RF},j} \right|\right)\nonumber \\
    = & \mathrm{log}_2\left(\left| 1 + \mathbf{f}_{\mathrm{RF},j}^{\rm H} \overline{\mathbf{B}}_{j} \mathbf{f}_{\mathrm{RF},j} \right|\right)\nonumber \\
    = & \mathrm{log}_2\left(\left| 1 + \mathbf{f}_{\mathrm{RF},j}^{*}\left(i\right)\sum_{m\neq i}^{M} \mathbf{f}_{\mathrm{RF},j}\left(m\right) \overline{\mathbf{B}}_{j} \left( i,m \right) \right. \right.\nonumber \\
    & \left.\left. \quad \quad \quad + \mathbf{f}_{\mathrm{RF},j}\left(i\right)\sum_{n\neq i}^{M} \mathbf{f}_{\mathrm{RF},j}^{*}\left(n\right) \overline{\mathbf{B}}_{j} \left( n,i \right) \right. \right.\nonumber \\
    & \left.\left. + \sum_{n\neq i}^{M} \mathbf{f}_{\mathrm{RF},j}^{*}\left(n\right) \sum_{m\neq i}^{M} \mathbf{f}_{\mathrm{RF},j}\left(m\right) \overline{\mathbf{B}}_{j} \left( n,m \right) + \frac{1}{M}\overline{\mathbf{B}}_{j} \left( i,i \right) \right| \right) \nonumber \\
    = & \mathrm{log}_2\left(\left| 1 + \mathfrak{R}\left\{ e^{-j\theta_{i,j}}\sum_{m\neq i}^{M} e^{j\theta_{m,j}} \overline{\mathbf{B}}_{j} \left( i,m \right)\right\}  \right. \right.\nonumber \\
    & \left.\left. + \sum_{n\neq i}^{M} e^{-j\theta_{n,j}} \sum_{m\neq i}^{M} e^{j\theta_{m,j}} \overline{\mathbf{B}}_{j} \left( n,m \right) + \frac{1}{M}\overline{\mathbf{B}}_{j} \left( i,i \right) \right| \right),
    \end{align}
    where $\overline{\mathbf{B}}_{j} = \frac{P_t }{\sigma^2 N_s}\mathbf{B}_{j} $.
    Therefore, for maximizing (\ref{decomp_1}), we update $\mathbf{f}_{\mathrm{RF},j}\left(i\right)$ for $\forall i \in \mathcal{S}_j$ as
    \begin{align}\label{upd_RF}
    \mathbf{f}_{\mathrm{RF},j}\left(i\right) = \frac{1}{\sqrt{M}} \frac{\sum_{m\neq i}^{M} e^{j\theta_{m,j}} \overline{\mathbf{B}}_{j} \left( i,m \right)}{\left| \sum_{m\neq i}^{M} e^{j\theta_{m,j}} \overline{\mathbf{B}}_{j} \left( i,m \right) \right|}.
    \end{align}
    Afterwards, we repeat the aforementioned joint connection and phase design method for determining the next element in $\mathbf{f}_{\mathrm{RF},j}$.
    By repeating the update of $\mathbf{f}_{\mathrm{RF},j}\left(i\right)$ for $\forall i \in \mathcal{S}_j$ via (\ref{upd_RF}) and the corresponding connection design, we are able to accomplish the design for all elements in $\mathbf{f}_{\mathrm{RF},j}$.
    When $c_{\rm H} = c_{\rm L} = \frac{M}{2}$, the design of the $j$-th sub-array in the twin-resolution dynamic sub-connected analog beamformer is deemed to be accomplished.
    Then we have $\overline{\mathbf{F}}_{\mathrm{RF},j}  = \begin{bmatrix} \overline{\mathbf{F}}_{\mathrm{RF},j-1}  & \mathbf{F}_{\mathrm{RF},j}\end{bmatrix}$.
%    Finally, we can acquire the twin-resolution dynamic sub-connected analog beamformer by designing the analog beamforming matrix column by column relying on SIC, which means that we design the twin-resolution dynamic sub-connected analog beamer sub-array by sub-array.
    Note that the proposed greedy method can also be invoked for deriving either an entirely high-resolution or low-resolution sub-connected analog beamformer by setting $N_{\rm L}=0$ or $N_{\rm H}=0$.
    The greedy method is summarized at a glance in Algorithm 1.
%    \textcolor{blue}{Note that the conventional designs for the phase shifters in conventional sub-connected structure \cite{JSAC_XGao_EnergyEfficient} and in twin-resolution fully-connected structure \cite{9219133} are not suitable for the proposed structure.
%Therefore, we proposed a corresponding greedy method, which are performed for all twin-resolution phase shifters that have not been designed in each sub-array.}

    \begin{algorithm}[h]
    \caption{Greedy Method Proposed for the $j$-th Sub-Array}
        \begin{algorithmic}[1]
            \REQUIRE ~~\\
            $\overline{\mathbf{F}}_{\mathrm{RF},j-1}$, $ \mathbf{H}_{{\rm eff},j}$, $\mathcal{Q}_{\rm H}$, $\mathcal{Q}_{\rm L}$, $\mathcal{S}_j = \left\{ 1,2,\cdots,M \right\}$, $c_{\rm L} = c_{\rm H} = 0$;
            \ENSURE
            \STATE Initialize $\mathbf{f}_{\mathrm{RF},1}$ as $\mathbf{f}_{\mathrm{RF},1} = \frac{1}{\sqrt{M}}e^{j\text{angle}\left( \mathbf{v}_1 \right)}$;
            \FOR{$i = 1 : M$}
            \STATE Design the connection between the $i_{\mathrm{L}}$-th or the $i_{\mathrm{H}}$-th TA and the corresponding phase shifter and set $\mathbf{f}_{\mathrm{RF},j}\left( i_{\rm L} \right) = \frac{1}{\sqrt{M}}e^{jQ_{\rm L}\left\{ q_{\rm L}\left( i_{\rm L} \right) \right\}}$ or $\mathbf{f}_{\mathrm{RF},j}\left( i_{\rm H} \right) = \frac{1}{\sqrt{M}}e^{jQ_{\rm H}\left\{ q_{\rm H}\left( i_{\rm H} \right) \right\}}$ according to the greedy method proposed in Section \ref{S4.1};
            \STATE Update $\mathcal{S}_j = \mathcal{S}_j\backslash \left\{ i_{\rm L} \right\}$ or $\mathcal{S}_j = \mathcal{S}_j\backslash \left\{ i_{\rm H} \right\}$;
            \STATE Update $c_{\rm L} = c_{\rm L} + 1$ or $c_{\rm H} = c_{\rm H} + 1$;
            \STATE Update residual elements in $\mathbf{f}_{\mathrm{RF},j}$ according to (\ref{decomp_1})-(\ref{upd_RF});
            \ENDFOR
            \label{code:recentEnd}
        \end{algorithmic}
    \end{algorithm}

    \subsection{CCM-based Method Proposed for Passive Beamforming Design}\label{S4.3}
    Following the design of the first $j$ sub-arrays, we proceed with the design of the passive beamforming matrix $\mathbf{\Phi}_{j}$ and update the effective channel $\mathbf{H}_{{\rm eff},j+1}$ for designing the remaining sub-arrays.
    We regard $\mathbf{G}\mathbf{\Phi}_j\mathbf{M} \overline{\mathbf{F}}_{\mathrm{RF},j}$ as the equivalent channel matrix \cite{chen2020hybrid} and formulate the passive beamforming optimization problem by
    \begin{subequations}\label{ObjSE_v5}
    \begin{align}
    \max_{ \mathbf{\Phi}_j } \quad & \mathrm{log}_2\left(\left| \mathbf{I}_{N_r} + \frac{P_t}{\sigma^2 j}  \mathbf{G}\mathbf{\Phi}_j\mathbf{M} \overline{\mathbf{F}}_{\mathrm{RF},j} \overline{\mathbf{F}}_{\mathrm{RF},j}^{\text{H}} \mathbf{M}^{\text{H}} \mathbf{\Phi}_j^{\text{H}} \mathbf{G}^{\text{H}} \right|\right) \\
    s.t. \quad & \mathbf{\Phi}_j = \text{diag}\left[ e^{j\phi_1},e^{j\phi_2},\cdots,e^{j\phi_{N_{\rm RIS}}} \right],
    \end{align}
    \end{subequations}
\newcounter{TempEqCnt} % 创建临时变量TempEqCnt
\setcounter{TempEqCnt}{\value{equation}} % 将当前公式序号 赋给TempEqCnt
\setcounter{equation}{20} % 当前公式序号变为x，x等于长公式应有的序号减1.
\begin{figure*}[ht]
    \begin{align}\label{Heff_trans3tp}
    &\left\| \mathbf{G}\mathbf{\Phi}_j\mathbf{M} \overline{\mathbf{F}}_{\mathrm{RF},j} \right\|_{\rm F}^2 \nonumber \\
    = & \mathrm{Tr}\left( \mathbf{G}\mathbf{\Phi}_j\mathbf{M} \overline{\mathbf{F}}_{\mathrm{RF},j} \overline{\mathbf{F}}_{\mathrm{RF},j}^{\text{H}} \mathbf{M}^{\text{H}} \mathbf{\Phi}_j^{\text{H}} \mathbf{G}^{\text{H}} \right) \nonumber \\
    = & \mathrm{Tr}\left( \begin{bmatrix} \pmb{\phi}_j^{\text{H}} & \mathbf{0} & \cdots & \mathbf{0} \\
    \mathbf{0} & \pmb{\phi}_j^{\text{H}} & \cdots & \mathbf{0} \\
    \mathbf{0} & \mathbf{0} & \cdots & \pmb{\phi}_j^{\text{H}}
    \end{bmatrix}
    \begin{bmatrix} \overline{\mathbf{M}}_{1}^{\text{H}}\mathbf{G}^{\text{H}} \\ \overline{\mathbf{M}}_{2}^{\text{H}}\mathbf{G}^{\text{H}}  \\ \vdots \\ \overline{\mathbf{M}}_{j}^{\text{H}}\mathbf{G}^{\text{H}}
    \end{bmatrix}
    \begin{bmatrix} \mathbf{G} \overline{\mathbf{M}}_{1} & \mathbf{G} \overline{\mathbf{M}}_{2} & \cdots & \mathbf{G} \overline{\mathbf{M}}_{j}
    \end{bmatrix}
    \begin{bmatrix} \pmb{\phi}_j & \mathbf{0} & \cdots & \mathbf{0} \\
    \mathbf{0} & \pmb{\phi}_j & \cdots & \mathbf{0} \\
    \mathbf{0} & \mathbf{0} & \cdots & \pmb{\phi}_j
    \end{bmatrix} \right)\nonumber \\
    = & \mathrm{Tr}\left(
    \begin{bmatrix} \pmb{\phi}_j^{\text{H}}\overline{\mathbf{M}}_{1}^{\text{H}}\mathbf{G}^{\text{H}} \mathbf{G} \overline{\mathbf{M}}_{1}\pmb{\phi}_j & \pmb{\phi}_j^{\text{H}}\overline{\mathbf{M}}_{1}^{\text{H}}\mathbf{G}^{\text{H}} \mathbf{G} \overline{\mathbf{M}}_{2}\pmb{\phi}_j & \cdots & \pmb{\phi}_j^{\text{H}}\overline{\mathbf{M}}_{1}^{\text{H}}\mathbf{G}^{\text{H}} \mathbf{G} \overline{\mathbf{M}}_{j}\pmb{\phi}_j\\ \pmb{\phi}_j^{\text{H}}\overline{\mathbf{M}}_{2}^{\text{H}}\mathbf{G}^{\text{H}} \mathbf{G} \overline{\mathbf{M}}_{1}\pmb{\phi}_j & \pmb{\phi}_j^{\text{H}}\overline{\mathbf{M}}_{2}^{\text{H}}\mathbf{G}^{\text{H}} \mathbf{G} \overline{\mathbf{M}}_{2}\pmb{\phi}_j & \cdots & \pmb{\phi}_j^{\text{H}}\overline{\mathbf{M}}_{2}^{\text{H}}\mathbf{G}^{\text{H}} \mathbf{G} \overline{\mathbf{M}}_{j}\pmb{\phi}_j\\
    \vdots & \vdots & \ddots & \vdots \\
    \pmb{\phi}_j^{\text{H}}\overline{\mathbf{M}}_{j}^{\text{H}}\mathbf{G}^{\text{H}} \mathbf{G} \overline{\mathbf{M}}_{1}\pmb{\phi}_j & \pmb{\phi}_j^{\text{H}}\overline{\mathbf{M}}_{j}^{\text{H}}\mathbf{G}^{\text{H}} \mathbf{G} \overline{\mathbf{M}}_{2}\pmb{\phi}_j & \cdots & \pmb{\phi}_j^{\text{H}}\overline{\mathbf{M}}_{j}^{\text{H}}\mathbf{G}^{\text{H}} \mathbf{G} \overline{\mathbf{M}}_{j}\pmb{\phi}_j
    \end{bmatrix} \right)\nonumber \\
    = & \pmb{\phi}_j^{\text{H}}\sum_{i=1}^{j}\overline{\mathbf{M}}_{i}^{\text{H}}\mathbf{G}^{\text{H}} \mathbf{G} \overline{\mathbf{M}}_{i}\pmb{\phi}_j.
    \end{align}
    \hrulefill
    \end{figure*}
    We further derive the upper bound of the OF (\ref{ObjSE_v5}a) as
    \setcounter{equation}{16} % 当前公式序号变为x，x等于长公式应有的序号减1.
%    \begin{align}\label{upb_Obj}
%    & \mathrm{log}_2\left(\left| \mathbf{I}_{N_r} + \frac{P_t }{\mathbf{\sigma}^2 j}  \mathbf{G}\mathbf{\Phi}_j\mathbf{M} \overline{\mathbf{F}}_{\mathrm{RF},j} \overline{\mathbf{F}}_{\mathrm{RF},j}^{\text{H}} \mathbf{M}^{\text{H}} \mathbf{\Phi}_j^{\text{H}} \mathbf{G}^{\text{H}} \right|\right) \nonumber \\
%    = & \mathrm{log}_2\left(\left| \mathbf{I}_{N_r} + \frac{P_t }{\sigma^2 j}  \mathbf{\Sigma}_{{\rm eff},j}^2 \right|\right) \nonumber \\
%    \overset{\left(a\right)}{\leq} & j\mathrm{log}_2\left(\left| 1 + \frac{P_t }{\sigma^2 j^2}  \left\|\mathbf{G}\mathbf{\Phi}_j\mathbf{M} \overline{\mathbf{F}}_{\mathrm{RF},j} \right\|_{\rm F}^2 \right|\right),
%    \end{align}
{\begin{align}\label{upb_Obj}
    %& \mathrm{log}_2\left(\left| \mathbf{I}_{N_r} + \frac{P_t }{\sigma^2 j}  \mathbf{\Sigma}_{{\rm eff},j}^2 \right|\right) \nonumber \\
    & \mathrm{log}_2\left(\left| \mathbf{I}_{N_r} + \frac{P_t}{\sigma^2 j}  \mathbf{G}\mathbf{\Phi}_j\mathbf{M} \overline{\mathbf{F}}_{\mathrm{RF},j} \overline{\mathbf{F}}_{\mathrm{RF},j}^{\text{H}} \mathbf{M}^{\text{H}} \mathbf{\Phi}_j^{\text{H}} \mathbf{G}^{\text{H}} \right|\right) \nonumber \\
    \overset{\left(a\right)}{\approx} & \sum_{i=1}^{j} \log_2\left(\left| 1 + \frac{P_t}{\sigma^2 j}  \mathbf{\Sigma}_{{\rm eff},j}^2\left( i,i \right) \right|\right) \nonumber \\
    \overset{\left(b\right)}{\leq} & j\log_2\left(\left| 1 + \frac{P_t }{\sigma^2 j^2} \sum_{i=1}^{j} \mathbf{\Sigma}_{{\rm eff},j}^2\left( i,i \right) \right|\right) \nonumber \\
    = & j\log_2\left(\left| 1 + \frac{P_t }{\sigma^2 j^2}  \left\|\mathbf{G}\mathbf{\Phi}_j\mathbf{M} \overline{\mathbf{F}}_{\mathrm{RF},j} \right\|_{\rm F}^2 \right|\right), \tag{17}
    \end{align}}
    where $\mathbf{\Sigma}_{{\rm eff},j}$ is a diagonal matrix with the diagonal elements being the singular values of $\mathbf{G}\mathbf{\Phi}_j\mathbf{M} \overline{\mathbf{F}}_{\mathrm{RF},j}$,
    {and $\left(a\right)$ holds for the approximation in \cite{wang2020joint} by relying on the truncated SVD of the equivalent channel matrix $\mathbf{G}\mathbf{\Phi}_j\mathbf{M} \overline{\mathbf{F}}_{\mathrm{RF},j}$, $\left(b\right)$ is obtained by adopting Jensen's inequality.}
    %and $\left(a\right)$ is obtained by adopting Jensen's inequality.
    We then maximize the upper bound of the bandwidth efficiency and transform the problem as
    \begin{subequations}\label{ObjSE_v6}
    \begin{align}
    \max_{ \mathbf{\Phi}_j } \quad & j\mathrm{log}_2\left(\left| 1 + \frac{P_t }{\sigma^2 j^2} \left\| \mathbf{G}\mathbf{\Phi}_j\mathbf{M} \overline{\mathbf{F}}_{\mathrm{RF},j} \right\|_{\rm F}^2 \right|\right) \\
    s.t. \quad & \mathbf{\Phi}_j = \text{diag}\left[ e^{j\phi_1},e^{j\phi_2},\cdots,e^{j\phi_{N_{\rm RIS}}} \right].
    \end{align}
    \end{subequations}
    However, the OF (\ref{ObjSE_v6}a) is still not in a tractable form.
    Thus, we substitute the block diagonal hybrid beamforming matrix (\ref{F}) into (\ref{ObjSE_v6}a).
    The item in $\left\| \cdot \right\|_{\rm F}^2$ can be equivalently transformed as
    \begin{align}\label{Heff_trans}
    & \mathbf{G}\mathbf{\Phi}_j\mathbf{M} \overline{\mathbf{F}}_{\mathrm{RF},j} \nonumber \\
    =& \mathbf{G}\mathbf{\Phi}_j
    \begin{bmatrix} \mathbf{M}_{1} & \mathbf{M}_{2} & \cdots & \mathbf{M}_{N_{\rm RF}}
    \end{bmatrix} \cdot \nonumber\\
    & \begin{bmatrix} \mathbf{f}_{\mathrm{RF},1} & \mathbf{0}_{M} & \cdots & \mathbf{0}_{M } \\
    \mathbf{0}_{M} & \mathbf{f}_{\mathrm{RF},2} & \cdots & \mathbf{0}_{M} \\
    \vdots & \vdots & \ddots & \vdots \\
    \mathbf{0}_{M} & \mathbf{0}_{M} & \cdots & \mathbf{f}_{\mathrm{RF},j} \\
    \mathbf{0}_{N_{\rm RF} - jM} & \mathbf{0}_{N_{\rm RF} - jM} & \cdots & \mathbf{0}_{N_{\rm RF} - jM}
    \end{bmatrix}\nonumber \\
    =& \mathbf{G}\mathbf{\Phi}_j \begin{bmatrix} \mathbf{M}_{1}\mathbf{f}_{\mathrm{RF},1} & \mathbf{M}_{2}\mathbf{f}_{\mathrm{RF},2} & \cdots & \mathbf{M}_{j}\mathbf{f}_{\mathrm{RF},j}
    \end{bmatrix},
    \end{align}
    where $\mathbf{M}_{i}$ contains the columns of $\mathbf{M}$ from the $M(i-1)+1$-st one to the $Mi$-th one, $\mathbf{0}_{a}$ denotes the all zero vector with the dimension of $a$.
    Then we have
    \begin{align}\label{Heff_trans3}
    & \mathbf{G}\mathbf{\Phi}_j\mathbf{M} \overline{\mathbf{F}}_{\mathrm{RF},j} \nonumber \\
    = & \mathbf{G}\mathbf{\Phi}_j \begin{bmatrix} \mathbf{M}_{1}\mathbf{f}_{\mathrm{RF},1} & \mathbf{M}_{2}\mathbf{f}_{\mathrm{RF},2} & \cdots & \mathbf{M}_{j}\mathbf{f}_{\mathrm{RF},j} \end{bmatrix}\nonumber \\
    = & \begin{bmatrix} \mathbf{G}\overline{\mathbf{M}}_{1}\pmb{\phi}_j & \mathbf{G}\overline{\mathbf{M}}_{2}\pmb{\phi}_j & \cdots & \mathbf{G}\overline{\mathbf{M}}_{j}\pmb{\phi}_j \end{bmatrix}\nonumber \\
    = & \begin{bmatrix} \mathbf{G} \overline{\mathbf{M}}_{1} & \mathbf{G} \overline{\mathbf{M}}_{2} & \cdots & \mathbf{G} \overline{\mathbf{M}}_{j}
    \end{bmatrix}
    \begin{bmatrix} \pmb{\phi}_j & \mathbf{0} & \cdots & \mathbf{0} \\
    \mathbf{0} & \pmb{\phi}_j & \cdots & \mathbf{0} \\
    \mathbf{0} & \mathbf{0} & \cdots & \pmb{\phi}_j
    \end{bmatrix},
    \end{align}
    where $\pmb{\phi}_j = \begin{bmatrix} e^{j\phi_1} & e^{j\phi_2} & \cdots & e^{j\phi_{N_{\rm RIS}}} \end{bmatrix}^{\text{T}}$, $\overline{\mathbf{M}}_{i}$ is a diagonal matrix with the $\left( k,k \right)$-th element being the $k$-th element of $\mathbf{M}_{i}\mathbf{f}_{\mathrm{RF},i}$.
    Then we can reformulate $\left\| \mathbf{G}\mathbf{\Phi}\mathbf{M} \overline{\mathbf{F}}_{\mathrm{RF},j} \right\|_{\rm F}^2$ as (\ref{Heff_trans3tp}) shown on the top of this page.
    Therefore, the problem (\ref{ObjSE_v6}) can be rewritten into a tractable form as
    \setcounter{equation}{21} % 当前公式序号变为x，x等于长公式应有的序号减1.
    \begin{subequations}\label{ObjSE_v7}
    \begin{align}
    \max_{ \pmb{\phi}_j } \quad & j\mathrm{log}_2\left(\left| 1 + \frac{P_t }{\sigma^2 j^2} \pmb{\phi}_j^{\text{H}}\mathbf{R}\pmb{\phi}_j \right| \right) \\
    s.t. \quad & \pmb{\phi}_j = \begin{bmatrix} e^{j\phi_1} & e^{j\phi_2} & \cdots & e^{j\phi_{N_{\rm RIS}}} \end{bmatrix}^{\text{T}},
    \end{align}
    \end{subequations}
    where $\mathbf{R} = \sum_{i=1}^{j}\overline{\mathbf{M}}_{i}^{\text{H}}\mathbf{G}^{\text{H}} \mathbf{G} \overline{\mathbf{M}}_{i}$.
    Note that the problem (\ref{ObjSE_v7}) has a classical form, hence we adopt the low-complexity CCM-based method for solving it \cite{9090356,9110849,wang2020joint}.
    Specifically, the search space of the problem (\ref{ObjSE_v7}) can be regarded as the product of $N_{\rm RIS}$ complex circles, each of which is $\mathcal{C} \overset{\Delta}{=}\left\{ u \in \mathbb{C}: u^{\text{H}}u = 1 \right\}$.
    The product of such $N_{\rm RIS}$ complex circles is a sub-manifold of $\mathbb{C}^{N_{\rm RIS}}$, which is regarded as CCM and defined by
    \begin{align}\label{M}
    \mathcal{M} = \mathcal{C}^{N_{\rm RIS}} \overset{\Delta}{=} \left\{ \mathbf{u} \in \mathbb{C}^{N_{\rm RIS}}: \left| u_i \right|=1, i=1,2,\cdots,M \right\},
    \end{align}
    where $u_i$ denotes the $i$-th element of $\mathbf{u}$.
    The main steps of the CCM-based method consist of three steps during the $t$-th iteration:
    \begin{enumerate}
    \item {Riemannian gradients:}

    The Riemannian gradient has the closed-form expression of \cite{absil2009optimization}
    \begin{align}\label{T_space}
    \mathcal{T}_{\pmb{\phi}_j^t}\mathcal{M} = \left\{ \mathbf{z} \in \mathbb{C}^{N_{\rm RIS}}: \mathfrak{R}\left\{ \mathbf{z} \circ\left( \pmb{\phi}_j^t \right)^{*} \right\} = \mathbf{0}\right\}.
    \end{align}
    The Riemannian gradient of the OF $f\left( \pmb{\phi}_j \right) = j\mathrm{log}_2\left(\left| 1 + \frac{P_t }{\sigma^2 j^2} \pmb{\phi}_j^{\text{H}}\mathbf{R}\pmb{\phi}_j \right| \right)$ at the point $\pmb{\phi}_j^t$ is a tangent vector $\bigtriangledown_{\mathcal{M}}f\left( \pmb{\phi}_j^t \right)$ given by \cite{absil2009optimization}
    \begin{align}\label{Riemannian_g}
    \bigtriangledown_{\mathcal{M}}f\left( \pmb{\phi}_j^t \right) & = \mathrm{Proj}_{\mathcal{T}_{\pmb{\phi}_j^t}\mathcal{M}}\left( \bigtriangledown f\left( \pmb{\phi}_j^t \right) \right) \nonumber \\
    & = \bigtriangledown f\left( \pmb{\phi}_j^t \right) - \mathfrak{R}\left\{ \bigtriangledown f\left( \pmb{\phi}_j^t \right) \circ\left( \pmb{\phi}_j^t \right)^{*} \right\} \circ \pmb{\phi}_j^t,
    \end{align}
    where $\mathrm{Proj}_{\mathcal{T}_{\pmb{\phi}_j^t}\mathcal{M}}\left( \cdot \right)$ is the orthogonal projection operator in the tangent space, while the Euclidean gradient $\bigtriangledown f\left( \pmb{\phi}_j^t \right)$ is in the direction opposite to the gradient of $f\left( \pmb{\phi}_j^t \right)$ given by
    \begin{align}\label{Euclidean_g}
    \bigtriangledown f\left( \pmb{\phi}_j^t \right) = -\frac{1}{\mathrm{ln}2}\frac{\frac{2 P_t }{\sigma^2 j}\mathbf{R}\pmb{\phi}_j^t}{1 + \frac{P_t }{\sigma^2 j^2} \left( \pmb{\phi}_j^t \right)^{\rm H}\mathbf{R}\pmb{\phi}_j^t}.
    \end{align}
    \item {Update over the tangent space:}

    We update $\pmb{\phi}_j^t$ on the tangent space with a Armijo step size $\omega^t$ of \cite{absil2009optimization}
    \begin{align}\label{upd_phi}
    \overline{\pmb{\phi}_j}^t = \pmb{\phi}_j^t - \omega^t \bigtriangledown_{\mathcal{M}}f\left( \pmb{\phi}_j^t \right).
    \end{align}
    \item {Retraction operator:}

    In order to map $\overline{\pmb{\phi}_j}^t$ into the CCM $\mathcal{M}$, we use the retraction operator, which is given by \cite{absil2009optimization}
    \begin{align}\label{Retraction}
    \overline{\pmb{\phi}_j}^{t+1} = \overline{\pmb{\phi}_j}^{t} \circ \frac{1}{ \overline{\pmb{\phi}_j}_{\mathrm{abs}}^{t} },
    \end{align}
    where the $i$-th element in $\overline{\pmb{\phi}_j}_{\mathrm{abs}}^{t}$ is the absolute of the $i$-th element in $\overline{\pmb{\phi}_j}^{t}$.
    \end{enumerate}
    The CCM-base method can converge to a critical point of the problem (\ref{ObjSE_v7}) \cite{absil2009optimization}, when the error between the OF (\ref{ObjSE_v7}a) before and after the final iteration is less than a predefined criterion $\epsilon$.
\subsection{Extension to discrete RIS scenarios}\label{S4.3}

{When considering discrete phases for the RIS elements, we adopt the common method of addressing the non-convex constraint of discrete space relying on approximation projection \cite{2002.03744v2}.
Specifically, we define the phase set, where the phases of RIS elements are chosen from the set $\mathcal{Q}_{\mathrm{RIS}}$ associated with $B_{\mathrm{RIS}}$ bits.
Following the proximity principle, we can project the resultant $\mathbf{\Phi}_j$ after designing the $j$-th sub-array to the elements in $\mathcal{Q}_{\mathrm{RIS}}$ relying on approximation projection given by
\begin{align}\label{projection}
\angle{\mathbf{\Phi}_j^{\rm dis}\left( i,i \right)} = \arg\min_{\psi \in \mathcal{Q}_{\mathrm{RIS}}} & \left| \angle{\mathbf{\Phi}_j\left( i,i \right)} - \psi \right|, \nonumber \\
&  i=1,2,\cdots,N_{\rm RIS},
\end{align}
where $\psi$ denotes the discrete phases in $\mathcal{Q}_{\mathrm{RIS}}$.
Afterwards, we substitute $\mathbf{\Phi}_j^{\rm dis}$ into Problem (\ref{ObjSE_v3_HB}) for the subsequent design steps.
}

\begin{algorithm}[h]
\caption{Overall Algorithm for Joint Hybrid and Passive Beamforming Design}
    \begin{algorithmic}[1]
        \REQUIRE ~~\\
        $\mathbf{G}$, $\mathbf{M}$, random generalized $\mathbf{\Phi}$;
        \ENSURE
        \STATE Initialize $\mathbf{F}_{\mathrm{RF}}$ as $\mathbf{F}_{\mathrm{opt}}$;
        \FOR{$j = 1 : N_{\rm RF}$}
        \STATE Update the analog beamforming matrix $\mathbf{f}_{{\rm RF},j}$ according to Algorithm 1;
        \STATE Calculate $\mathbf{F}_{\mathrm{RF},j}$ according to (\ref{F});
        \STATE Calculate $\overline{\mathbf{F}}_{\mathrm{RF},j}  = \begin{bmatrix} \overline{\mathbf{F}}_{\mathrm{RF},j-1}  & \mathbf{F}_{\mathrm{RF},j}\end{bmatrix}$;
        \STATE Update the passive beamforming matrix $\mathbf{\Phi}_j$ relying on the CCM-based method until meeting the convergence criterion $\epsilon$;
        \ENDFOR
        \STATE Calculate $\mathbf{F}_{\mathrm{RF}} = \overline{\mathbf{F}}_{\mathrm{RF},{N_{\rm RF}}}$;
        \STATE Calculate $\mathbf{\Phi} = \mathbf{\Phi}_{N_{\rm RF}}$;
        \STATE Calculate $\mathbf{F}_{\mathrm{BB}} = \left( \mathbf{F}_{\mathrm{RF}}^{\rm H}\mathbf{F}_{\mathrm{RF}} \right)^{-\frac{1}{2}}\mathbf{V}_{\rm eff}$;
        \label{code:recentEnd3}
    \end{algorithmic}
\end{algorithm}
    \subsection{Overall Algorithm and Complexity Analysis}\label{S4.4}
    The overall SIC-based joint hybrid and passive beamforming design is summarized in Algorithm 2.
    We update one of the variables, while the others are fixed when designing each sub-array.
    Once all sub-arrays have been designed, our joint design is accomplished.

    \textit{{Remark 1}}: In our SIC-based joint design method, we use the analog and  passive beamforming matrices previously obtained for designing the remaining columns, which is based on SIC.
    Furthermore, according to the greedy method, for designing the remaining elements in a sub-array, we update the elements according to (\ref{upd_RF}) that have not yet been designed in the analog beamforming matrix.
    Therefore, the quantization error of all previously designed phases is taken into consideration when we design the remaining phases and update the RIS elements.

    {\textit{{Remark 2}}: Given the approximation procedures used during the problem transformation, the SIC scheme and the specific properties of the proposed greedy method employed for configuring the twin-resolution phase shifters of each sub-array, our proposed joint hybrid and passive beamforming design method delivers feasible solutions, rather than globally or locally optimal solutions.}
    Nevertheless, they can achieve near-optimal spectral performance according to the simulation results of Section \ref{S6}.

%    \textcolor{blue}{\textit{{Remark 3}}: Since perfect CSI is impractical to obtain, we also show an idea for channel estimation. Specifically, inspired by the channel estimation procedure in \cite{zhou2021channel}, where the angles are assumed to be unchanged for multiple coherence blocks, while the channel gains change from block to block. Therefore, in the first coherence block, the full CSI information can be estimated by an atomic norm minimization algorithm for RIS-assisted MIMO systems \cite{9398559}. Afterwards, the channel gains can be estimated in the remaining coherence blocks via orthogonal matching pursuit (OMP)-based method relying on the problem transformation in \cite{9103231}.}

    The computational complexity is analyzed as follows.
    For our hybrid beamforming design, the SVD of calculating $\mathbf{v}_j$ for $j = 1,2,\cdots,N_{\rm RF}$ has the computational complexity order of $\mathcal{O}\left( M^3N_{\rm RF}\right)$.
    The update of $\mathbf{f}_{\mathrm{RF},j}\left(i\right)$ for $i = 1,2,\cdots,M$, $j = 1,2,\cdots,N_{\rm RF}$ in (\ref{upd_RF}) has the computational complexity order of $\mathcal{O}\left( M - 1\right)$, which has to be repeated $ \frac{M\left( M-1 \right)}{2} N_{\rm RF} $ times for completing the analog beamforming matrix design.
    Thus, the computational complexity order of the greedy method is $\mathcal{O}\left( M^3N_{\rm RF} + \frac{M\left( M-1 \right)^2}{2} N_{\rm RF} \right)$.
    For our passive beamforming design, much of the complexity is associated with calculating the Riemannian gradient, which can be accomplished by calculating $ \mathbf{R}\pmb{\phi}_j^t $ during the $t$-th iteration.
    The complexity order of the CCM-based passive beamforming design is $\mathcal{O}\left( N_{\rm RIS}N_{\rm RF}I_1\right)$, where $I_1$ is the number of iterations required for convergence.
    Therefore, the complexity order of the overall algorithm is $\mathcal{O}\left(  M^3N_{\rm RF} + \frac{M\left( M-1 \right)^2}{2} N_{\rm RF} + N_{\rm RIS}N_{\rm RF}I_1\right)$.
    \section{Numerical Results}\label{S5}
    In this section, we provide simulation results for characterizing the proposed joint hybrid and passive beamforming design of RIS-assisted mmWave systems relying on twin-resolution dynamic sub-connected hybrid structures.
 	\begin{figure}[t]
		\center{\includegraphics[width=0.45\textwidth]{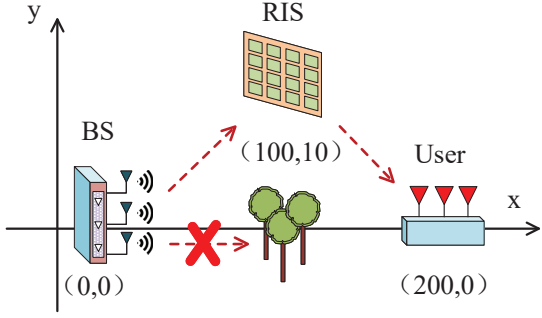}}
		\caption{Illustration of our simulation scenario.}
		\label{Simulations}
	\end{figure}

    Unless otherwise stated, the number of TAs in our simulations is set to $N_t = 8 \times 8 = 64$, the number of RF chains is $N_{\rm RF} = 4$, the number of complex-gain propagation paths is $L = P =4$, while the number of data streams is set to $N_s = 4$.
    The azimuth AoA $\phi^{\mathrm{r}}_{\ell}$ and AoD $\phi^{\mathrm{t}}_{\ell}$ are uniformly distributed in the interval $\left[ -\pi,\pi \right)$ and the elevation AoA $\theta^{\mathrm{r}}_{\ell}$ and AoA $\theta^{\mathrm{t}}_{\ell}$ are uniformly distributed in the interval $\left[ -\frac{\pi}{2},\frac{\pi}{2} \right)$ \cite{TWC_AAhmed_LimitedHybridPrecoding}.
    As illustrated in Fig. \ref{Simulations}, the coordinates of the BS, the RIS and the user are $\left( x_{\rm BS}, y_{\rm BS} \right)=\left( 0, 0 \right)$m, $\left( x_{\rm RIS}, y_{\rm RIS} \right)=\left( 100, 10 \right)$m, $\left( x_{\rm user}, y_{\rm user} \right)=\left( 200, 0 \right)$m, respectively.
    The distances between the BS and the RIS and that between the RIS and the user are denoted by $d_{\rm BS-RIS}$ and $d_{\rm RIS-user}$, respectively.
    {We calculate the system's path loss according to the theoretical free-space distance-dependent path-loss model for investigating the bandwidth efficiency of our proposed structure and beamforming methods \cite{8811733,zhang2010channel}.}
    The path-loss in the BS-RIS link and RIS-user link is then calculated by $A_{\rm BR} = -30 - 22.0\log d_{\rm BS-RIS}$ and $A_{\rm RU} = -30 - 22.0\log d_{\rm RIS-user}$, respectively \cite{8811733}.
    The carrier frequency is $28$GHz and the bandwidth is $B = 251.1886$MHz \cite{wang2020joint}.
    The noise power is calculated by $\sigma^2 = -174 + 10\log_{10}B = -90$dBm.
    In our simulations, the higher resolution of the phase shifter is set to $B_{\rm H} = 4$bits and the lower resolution is set to $B_{\rm L} = 1$bit.
    We set the error threshold of passive beamforming to $\epsilon = 10^{-6}$.
    The key parameters are listed in Table II.

    \begin{table}\label{parameter1}
    \centering
    \caption{Key Parameters}
    \begin{tabular*}{7.5cm}{l r}
    \hline
    Parameter & Value \\
    \hline
    Number of paths, $L$, $P$ & 4 \\
    Distances, $d_{\rm BS-RIS}$, $d_{\rm RIS-user}$  & 100.4988m \\
    Path Loss, $A_{\rm BR}$, $A_{\rm RU}$ & -74.0475dB \\
    Carrier frequency & 28GHz \\
    Bandwidth, $B$ & 251.1886MHz \cite{wang2020joint} \\
    Noise power, $\sigma^2$ & -90dBm \\
    \hline
    \end{tabular*}
    \end{table}
 	\begin{figure}[t]
		\center{\includegraphics[width=0.45\textwidth]{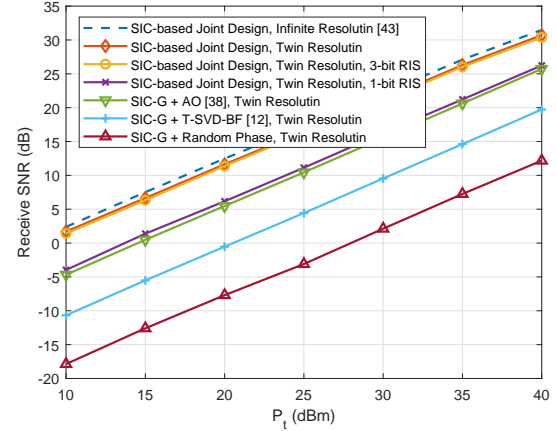}}
		\caption{Receive SNR vs. transmit power.}
		\label{RSNR_vs_rho}
	\end{figure}

 	\begin{figure}[t]
		\center{\includegraphics[width=0.45\textwidth]{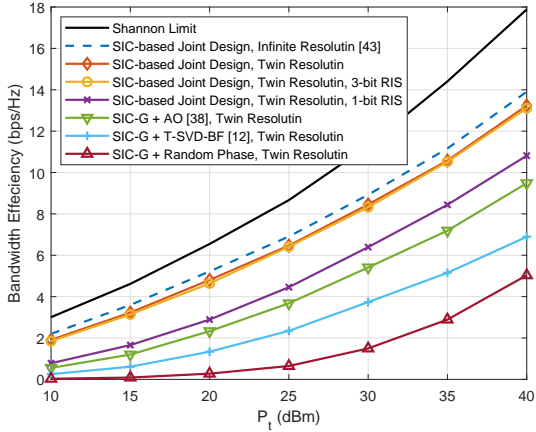}}
		\caption{Bandwidth Efficiency vs. transmit power.}
		\label{SE_vs_Ps}
	\end{figure}
 	\begin{figure}[t]
		\center{\includegraphics[width=0.45\textwidth]{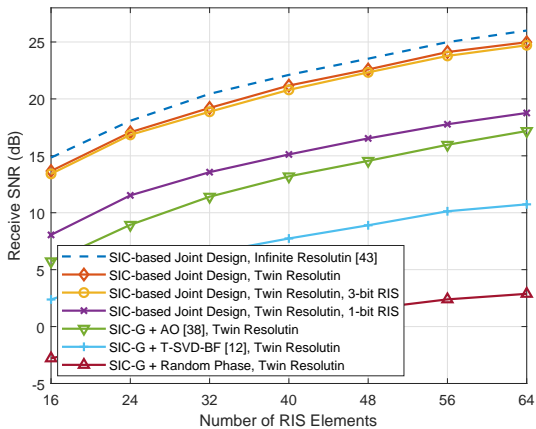}}
		\caption{Receive SNR vs. number of RIS elements.}
		\label{RSNR_vs_NRIS}
	\end{figure}

    \subsection{Receive SNR and bandwidth efficiency comparison}\label{S6.1}
    In Fig. \ref{RSNR_vs_rho}, we show the relationship between the receive SNR defined by $\frac{P_t  \left\|\mathbf{G}\mathbf{\Phi}\mathbf{M}\mathbf{F}_{\rm RF}\mathbf{F}_{\rm BB}\right\|_{\rm F}^2}{\sigma^2 N_s}$ and the transmit power $P_t$.
    Since the hybrid beamforming matrix $\mathbf{F}_{\rm RF}\mathbf{F}_{\rm BB}$ and RIS matrix $\mathbf{\Phi}$ are optimized for a given transmit power constraint $P_t$, the effective channel (including RIS reflection) and the receive SNR are different, when we adopt different hybrid/passive beamforming design methods.
    {We consider 7 curves for comparison, including:}
    1) \textit{SIC-based Joint Design, Infinite Resolution \cite{JSAC_XGao_EnergyEfficient}}: we adopt the SIC-based joint design method for hybrid beamforming design, where the greedy method is replaced by the algorithm of \cite{JSAC_XGao_EnergyEfficient} considering the infinite resolution of phase shifters;
    %2) SIC-based Joint Design, High Resolution: we adopt the proposed method for jointly designing the hybrid and passive beamforming matrices.
%    In each sub-array, there exist entirely high-resolution phase shifters.
    2) \textit{SIC-based Joint Design, Twin Resolution}: we adopt the method advocated for the proposed structure;
    {3) \textit{SIC-based Joint Design, Twin Resolution, 3-bit RIS}: we adopt the method advocated and the resolution of the RIS elements is $B_{\rm RIS} = 3$;
    4) \textit{SIC-based Joint Design, Twin Resolution, 1-bit RIS}: This method is the same as 3) except that $B_{\rm RIS} = 1$;}
    5) \textit{SIC-G + AO \cite{9110912}, Twin Resolution}: we adopt the proposed SIC-based greedy beamforming method for the hybrid beamforming design and the AO-based method of \cite{9110912} for passive beamforming design, where twin-resolution of phase shifters are considered;
    6) \textit{SIC-G + T-SVD-BF \cite{wang2020joint}}, Twin Resolution: we adopt the proposed SIC-based greedy beamforming method for hybrid beamforming design and the truncated SVD beamforming (T-SVD-BF) method of \cite{wang2020joint} for passive beamforming design, where the phase shifters are of twin-resolution;
    7) \textit{SIC-G + Random Phase, Twin Resolution}: the phases of RIS elements are randomly set for comparison.
    Fig. \ref{RSNR_vs_rho} demonstrates that our structure conceived with our proposed SIC-based joint design achieves higher receive SNR than the other methods.
    {Fig. 5 also illustrates that when the resolution of the RIS elements is $B_{\rm RIS}=3$bits, the proposed quantization method approaches the bandwidth efficiency of the infinite-resolution RIS scenarios.}
%Even when the resolution of RIS element is $B_{\rm RIS}=1$bit, our proposed method together with the proposed structure has higher bandwidth efficiency compared to the conventional methods.}
    Moreover, there is a 20dB receive SNR gain, when the RIS elements are optimized.
    With the aid of the proposed joint hybrid and passive beamforming method, we can improve the receive SNR by 20dB.

    In Fig. \ref{SE_vs_Ps}, we show the bandwidth efficiency vs. the transmit power $P_t$ \footnote{Using the transmit power is unconventional. This quantity is however beneficial for the problem considered, because the proposed method increases the receive SNR for given transmit power.}.
    Observe that our solution outperforms the benchmarks, because it relies on a joint hybrid and passive beamforming design.
%    Moreover, our transformed upper bound maximization problem (\ref{ObjSE_v6}) relies on relaxation procedures to a lesser extent than the passive beamforming design of \cite{9110912}.
    %%Even when entirely low-resolution (1-bit) phase shifters are considered in the analog beamformer, our proposed method can achieve satisfying bandwidth efficiency performance.
%    In Fig. \ref{SE_vs_Ps}, we can also observe that the proposed structure harnessing the SIC-based joint design achieves near-optimal performance, outperforming the conventional sub-connected hybrid beamforming structure of \cite{JSAC_XGao_EnergyEfficient}.
    We also add \textit{Shannon Limit} to Fig. \ref{SE_vs_Ps}, which is obtained by the Shannon limit defined in \cite{cho2010mimo} with the effective channel obtained by the proposed joint hybrid and passive beamforming matrix.
    Nevertheless, the hybrid beamforming structure, finite/low-resolution phase shifters and the approximations adopted during the problem transformation inevitably erode the system's performance.
    Therefore, our proposed method lags behind than the Shannon limit.

    Fig. \ref{RSNR_vs_NRIS} illustrates the relationship between the receive SNR and the number $N_{\rm RIS}$ of RIS elements.
    As shown in Fig. \ref{RSNR_vs_NRIS}, the receive SNR increases as $N_{\rm RIS}$ increases, which is due to the fact that having more RIS elements leads to higher beamforming gains.
    Furthermore, when $N_{\rm RIS}$ is increased, the receive SNR tends to saturate, owing to the maximum SNR given by $\frac{P_t}{\sigma^2}$.

 	\begin{figure}[t]
		\center{\includegraphics[width=0.45\textwidth]{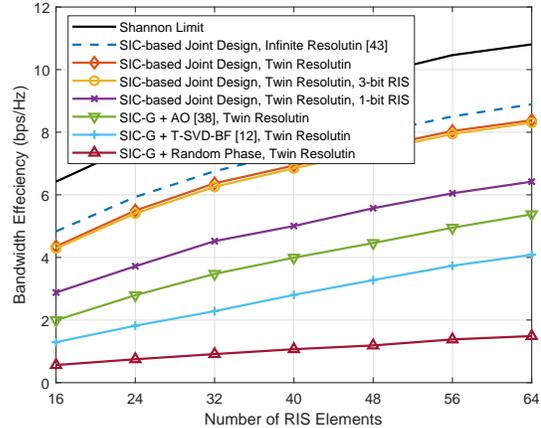}}
		\caption{Bandwidth efficiency vs. number of RIS elements.}
		\label{SE_vs_NRIS}
	\end{figure}

 	\begin{figure}[t]
		\center{\includegraphics[width=0.45\textwidth]{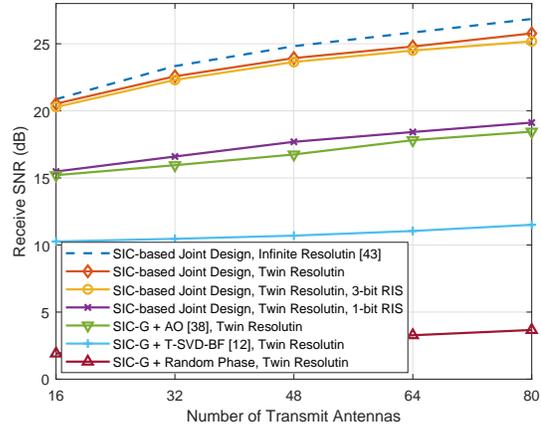}}
		\caption{Receive SNR vs. number of transmit antennas.}
		\label{RSNR_vs_Nt}
	\end{figure}
 	\begin{figure}[t]
		\center{\includegraphics[width=0.45\textwidth]{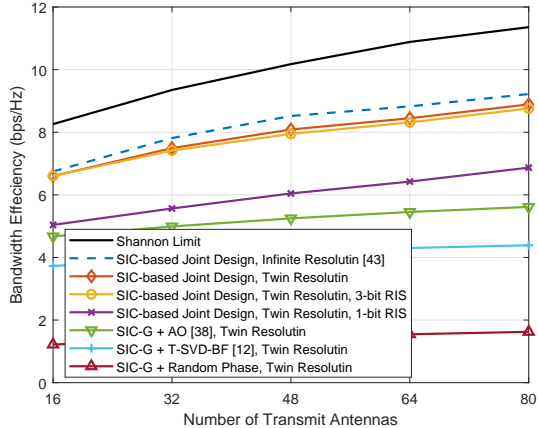}}
		\caption{Bandwidth efficiency vs. number of transmit antennas.}
		\label{SE_vs_TA}
	\end{figure}

    Fig. \ref{SE_vs_NRIS} shows the relationship between the bandwidth efficiency and the number $N_{\rm RIS}$ of RIS elements.
    The relative relationship among the 6 curves is reminiscent of those in Fig. \ref{SE_vs_Ps}.
    Moreover, we observe that as $N_{\rm RIS}$ increases from $16$ to $64$, the system's bandwidth efficiency increases in line with the Shannon limit, because more RIS elements increase the beamforming gain of the effective channel $\mathbf{G}\mathbf{\Phi}\mathbf{M}$.

    The relationship between the receive SNR and the number $N_t$ of TAs is shown in Fig. \ref{RSNR_vs_Nt}.
    As $N_t$ increases from $16$ to $80$, the number of antennas in each sub-array $M$ increases from $4$ to $20$, which improves the beamforming gain.
    Similar to the above analysis, the receive SNR cannot be larger than $\frac{P_t}{\sigma^2}$.

    We show the relationship between the bandwidth efficiency and $N_t$ in Fig. \ref{SE_vs_TA}.
    Since more TAs lead to higher beamforming gain for the effective channel $\mathbf{G}\mathbf{\Phi}\mathbf{M}$, both the Shannon limit and the bandwidth efficiency increase upon increasing $N_t$.
    \subsection{Bandwidth efficiency vs. energy efficiency trade-off}\label{S6.2}
    \begin{table}\label{parameter2}
    \centering
    \caption{Power Consumption of Devices}
    \begin{tabular*}{8.5cm}{l r}
    \hline
    Device & Power Consumption \\
    \hline
    Baseband processor, $P_{\mathrm{BB}}$  & 200 mW \cite{7997065} \\
    Each RF chain, $P_{\mathrm{RF}}$  & 300 mW \cite{7997065} \\
    Each high-resolution phase shifter, $P_{\rm H}$ (4 bits)  & 52 mW \cite{mendez2016hybrid} \\
    Each low-resolution phase shifter, $P_{\rm L}$ (1 bit) & 10 mW \cite{mendez2016hybrid} \\
    Each switch, $P_{\mathrm{SW}}$ & 1mW \cite{9219133} \\
    Each RIS element, $P_{\mathrm{RIS}}$ & 1mW \cite{8741198} \\
    \hline
    \end{tabular*}
    \end{table}
 	\begin{figure}[t]
		\center{\includegraphics[width=0.45\textwidth]{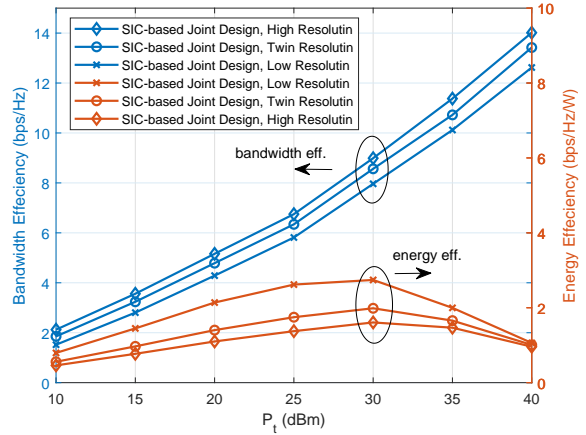}}
		\caption{Bandwidth efficiency $\&$ energy efficiency trade-off vs. transmit power.}
		\label{EE_vs_Ps}
	\end{figure}
    Fig. \ref{EE_vs_Ps} illustrates the bandwidth efficiency and energy efficiency vs. the transmit power $P_t$.
    Note that our proposed method can be used for designing the hybrid beamforming matrix entirely relying on high-resolution or low-resolution phase shifters by setting $N_{\rm L}=0$ or $N_{\rm H}=0$ in the proposed method.
    Three scenarios are considered, including
    1) \textit{SIC-based Joint Design, High Resolution}: we adopt the proposed method for jointly designing the hybrid and passive beamforming matrices.
        In each sub-array, only high-resolution phase shifters are used;
    2) \textit{SIC-based Joint Design, Twin Resolution}: we adopt the proposed method for the advocated structure;
    3) \textit{SIC-based Joint Design, Low Resolution}: this scenario is similar to 1) except that only low-resolution phase shifters are adopted in each sub-array.
{Although controlling all switches will lead to some energy loss \cite{9026753}, the main power dissipation of the dynamically configurable switching network advocated is due to the total power of the switches enabled during the transmission \cite{8903536,7370753}. Hence the power consumption of the switches is given by $N_{t} P_{\mathrm{SW}}$.}
    The energy efficiency of the proposed RIS-assisted mmWave MIMO communication systems relying on twin-resolution dynamic sub-connected hybrid structures is calculated by $\mathrm{EE}_{\rm Twin} = \frac{R}{ P_{\rm Twin} }$,
%    \begin{align}\label{EE}
%    \mathrm{EE}_{\rm Twin} = \frac{R}{ P_{\rm Twin} },
%	\end{align}
    where $P_{\rm Twin}$ is the total power consumption of the systems, which is calculated by $P_{\rm Twin} = P_t + P_{\mathrm{BB}} + N_{\rm RF} P_{\mathrm{RF}} + N_{\rm RIS}P_{\mathrm{RIS}}  + \frac{N_t}{2}P_{\rm H} + \frac{N_t}{2}P_{\rm L} + N_{t} P_{\mathrm{SW}}$,
%    \begin{align}\label{P_total}
%    P_{\rm Twin} = P_t + P_{\mathrm{BB}} + N_{\rm RF} P_{\mathrm{RF}} + N_{\rm RIS}P_{\mathrm{RIS}}  + \frac{N_t}{2}P_{\rm H} + \frac{N_t}{2}P_{\rm L} + + N_{t} P_{\mathrm{SW}},
%	\end{align}
    and $P_{\mathrm{BB}}$ is the power consumption of the baseband processor, and $P_{\mathrm{RF}}$ denotes the power consumption of each RF chain.
    The symbols $P_{\rm H}$ and $ P_{\rm L} $ are the power consumption of a high-resolution and of a low-resolution phase shifter, respectively.
    The symbol $P_{\mathrm{SW}}$ represents the power consumption of each switch in the systems, which is used for connecting the phase shifters to the TAs.
    The symbol $P_{\mathrm{RIS}}$ stands for the power consumption of each RIS element.
    When considering the purely high- or solely low-resolution sub-connected analog beamformer, the energy calculation is slightly modified by $\mathrm{EE}_{o} = \frac{R}{ P_o }$,
%    \begin{align}\label{EE}
%    \mathrm{EE}_{o} = \frac{R}{ P_o },
%	\end{align}
    where $ o \in \left\{ \mathrm{H}, \mathrm{L} \right\}$, and $P_o$ is the total power consumption defined by $P_o = P_t + P_{\mathrm{BB}} + N_{\rm RF} P_{\mathrm{RF}} + N_{\rm RIS}P_{\mathrm{RIS}} + N_tP_{\rm o}$.
%    \begin{align}\label{P_total}
%    P_o = P_t + P_{\mathrm{BB}} + N_{\rm RF} P_{\mathrm{RF}} + N_{\rm RIS}P_{\mathrm{RIS}} + N_tP_{\rm o}.
%	\end{align}
    The power consumptions of the devices in the system are presented in Table III.
    We can draw the conclusion for Fig. \ref{EE_vs_Ps} that the twin-resolution dynamic sub-connected hybrid beamforming structure strikes a bandwidth efficiency vs. energy efficiency trade-off.
    Specifically, the RIS-assisted MIMO systems using twin-resolution dynamic sub-connected analog beamformers have a higher bandwidth efficiency than that of the entirely low-resolution sub-connected analog beamformer, while its energy efficiency is lower.
    By contrast, although the RIS-assisted system relying on the proposed structure has a lower bandwidth efficiency than the entirely high-resolution sub-connected hybrid structure, it is more energy-efficient.
    Observe in Fig. \ref{EE_vs_Ps} that the energy efficiency increases first and then decreases.
    This is because the energy efficiency is calculated as a ratio, where the bandwidth efficiency in the numerator is a logarithmic function of $P_t$, while the power consumption in the denominator is a linear function of $P_t$.
    Therefore, the numerator increases more rapidly than the denominator first, but slower beyond the maximum.
    The optimal transmit power $P_t$ for maximum EE is about 30dBm.
 	\begin{figure}[t]
		\center{\includegraphics[width=0.45\textwidth]{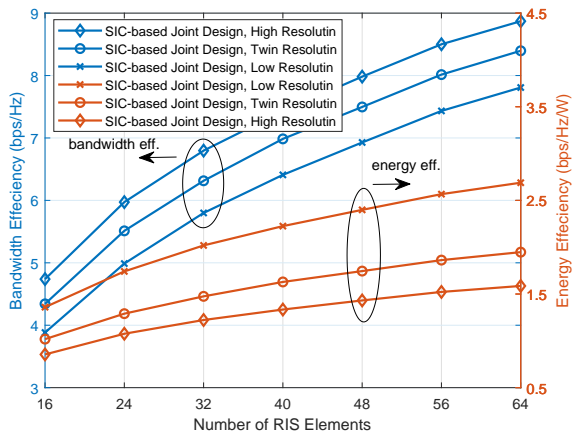}}
		\caption{Bandwidth efficiency $\&$ energy efficiency trade-off vs. number of RIS elements.}
		\label{EE_vs_NRIS}
	\end{figure}
 	\begin{figure}[t]
		\center{\includegraphics[width=0.45\textwidth]{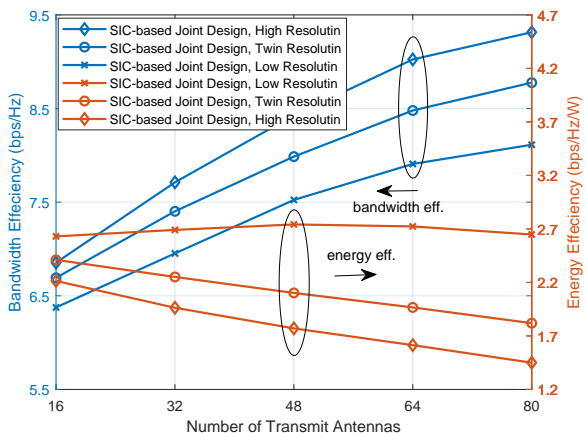}}
		\caption{Bandwidth efficiency $\&$ energy efficiency trade-off vs. number of transmit antennas.}
		\label{EE_vs_TA}
	\end{figure}
    Fig. \ref{EE_vs_NRIS} portrays both the bandwidth efficiency and energy efficiency versus the number of RIS elements.
    Similar to the trends in Fig. \ref{EE_vs_Ps}, our proposed structure strikes a trade-off between bandwidth efficiency and energy efficiency, when the number of RIS elements increases from $16$ to $64$.
    Furthermore, as the number of RIS elements increases, both the system's bandwidth efficiency and energy efficiency increases for all three cases, but gradually saturates.

    A bandwidth efficiency vs. energy efficiency trade-off can also be observed from Fig. \ref{EE_vs_TA}, as the number TAs increases from $16$ to $80$.
    We observe that the energy efficiency of the entirely low-resolution sub-connected hybrid structure based RIS-assisted MIMO systems is better than that of the twin-resolution dynamic sub-connected hybrid structure based ones and of the entirely high-resolution sub-connected hybrid structure based schemes.
    \section{Conclusion}\label{S6}
    In this paper, we proposed RIS-assisted millimeter-wave (mmWave) wireless communication systems relying on twin-resolution dynamic sub-connected hybrid beamforming structures.
    More explicitly, we conceived an energy-efficient and low-complxity dynamic sub-connected analog beamformer relying on twin-resolution phase shifters, which combines the advantages of the high spatial gain brought about by high-resolution phase shifters and the cost efficiency of low-resolution phase shifters.
    The RIS enhances the performance of communication systems by beneficially reflecting the multi-path signals.
    For jointly designing the hybrid beamforming and RIS elements, a SIC-based joint design method was proposed.
    Specifically, we decomposed the bandwidth efficiency maximization problem into a series of sub-rate maximization problems, where the sub-array of phase shifters and RIS elements were jointly optimized.
    For our hybrid beamforming design, we proposed a greedy method for beneficially configuring the connections between the TAs of each sub-array and the phase shifters.
    For our passive beamforming design, we developed a CCM-based method for updating the RIS elements.
    Our simulation results demonstrated the superiority of the proposed structure and the proposed method over its counterparts.
    In conclusion, the proposed structure combined with the proposed method strikes an attractive trade-off between the bandwidth efficiency and power consumption.%\cite{9484683}

%    \cite{8443089}

    \bibliographystyle{IEEEtran}
	\bibliography{IEEEabrv,Refference}
\begin{IEEEbiography}[{\includegraphics[width=1in,height=1.25in,clip,keepaspectratio]{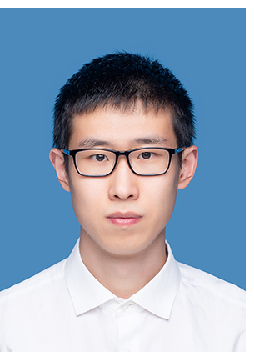}}]
		{Chenghao Feng} received the B.E. degree from the Beijing Institute of Technology, Beijing, China, in 2017, where he is currently pursuing the Ph.D. degree with the School of Information and Electronics. His current research interests include massive MIMO, mmWave/THz communications, energy-efficient communications, intelligent reflecting surface and networks.
\end{IEEEbiography}

\begin{IEEEbiography}[{\includegraphics[width=1in,height=1.25in,clip,keepaspectratio]{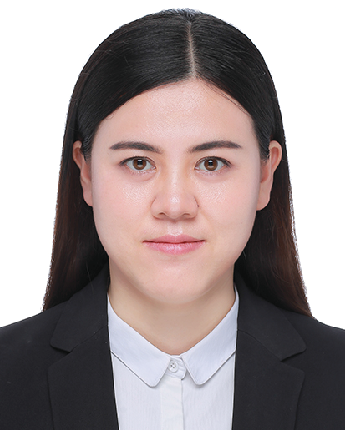}}]
		{Wenqian Shen} received the B.S. degree from Xi'an Jiaotong University, Shaanxi, China in 2013 and the Ph.D. degree from Tsinghua University, Beijing, China in 2018. She is currently an associate professor with the School of Information and Electronics, Beijing Institute of Technology, Beijing, China. Her research interests include massive MIMO and mmWave/THz communications. She has published several journal and conference papers in IEEE Transaction on Signal Processing, IEEE Transaction on Communications, IEEE Transaction on Vehicular Technology, IEEE ICC, etc. She has won the IEEE Best Paper Award at the IEEE ICC 2017.
\end{IEEEbiography}

	\begin{IEEEbiography}[{\includegraphics[width=1in,height=1.25in,clip,keepaspectratio]{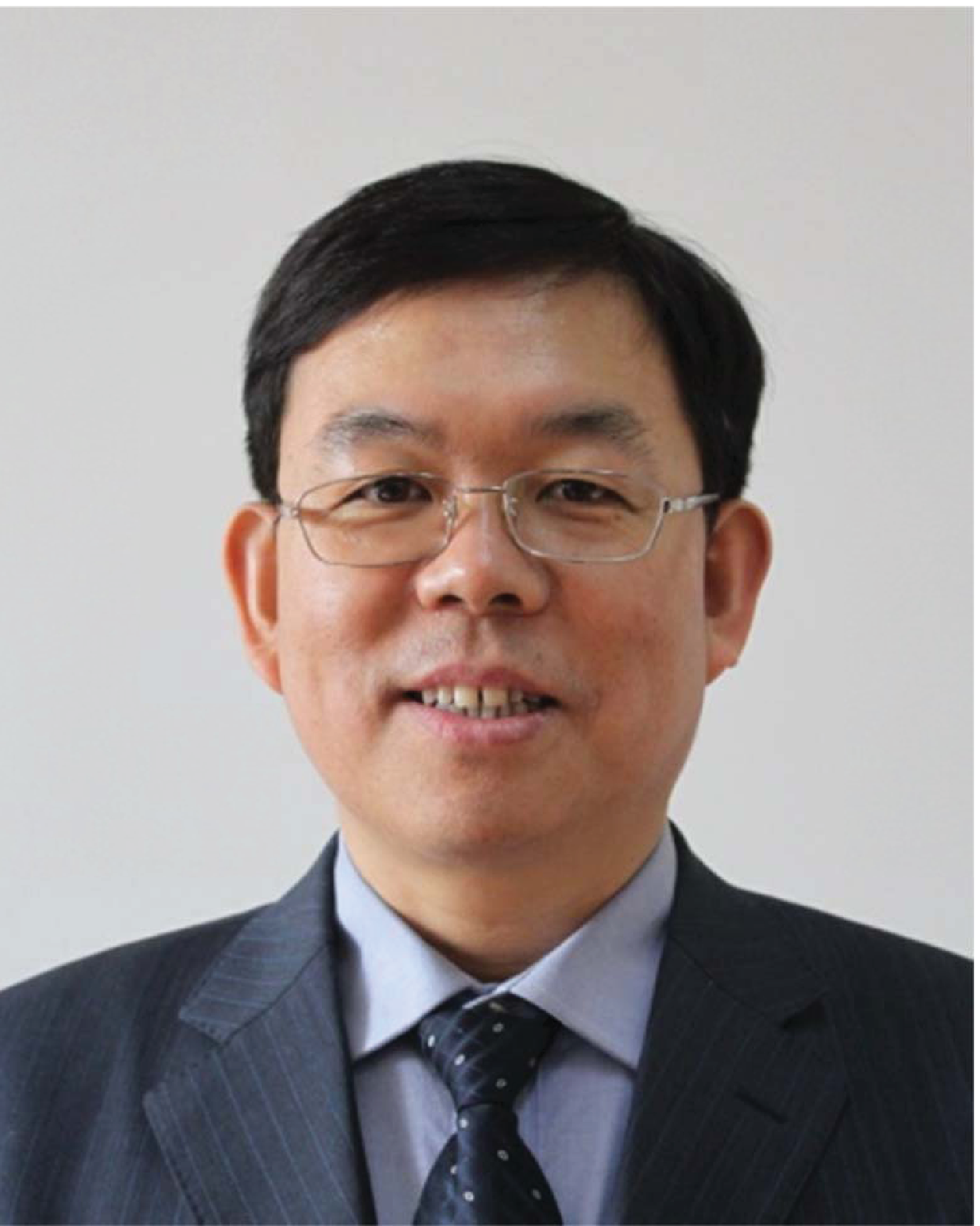}}]
		{Jianping An} (M'08) received the B.E. degree from Information Engineering University in 1987, and the M.S. and Ph.D. degrees from Beijing Institute of Technology, in 1992 and 1996, respectively. Since 1996, he has been with the School of Information and Electronics, Beijing Institute of Technology, where he now holds the post of Full Professor. From 2010 to 2011, he was a Visiting Professor at University of California, San Diego. He has published more than 150 journal and conference articles and holds (or co-holds) more than 50 patents. He has received various awards for his academic achievements and the resultant industrial influences, including the National Award for Scientific and Technological Progress of China (1997) and the Excellent Young Teacher Award by the China's Ministry of Education (2000). Since 2010, he has been serving as a Chief Reviewing Expert for the Information Technology Division, National Scientific Foundation of China. Prof. An's current research interest is focused on digital signal processing theory and algorithms for communication systems.
	\end{IEEEbiography}

\begin{IEEEbiography}[{\includegraphics[width=1in,height=1.25in,clip,keepaspectratio]{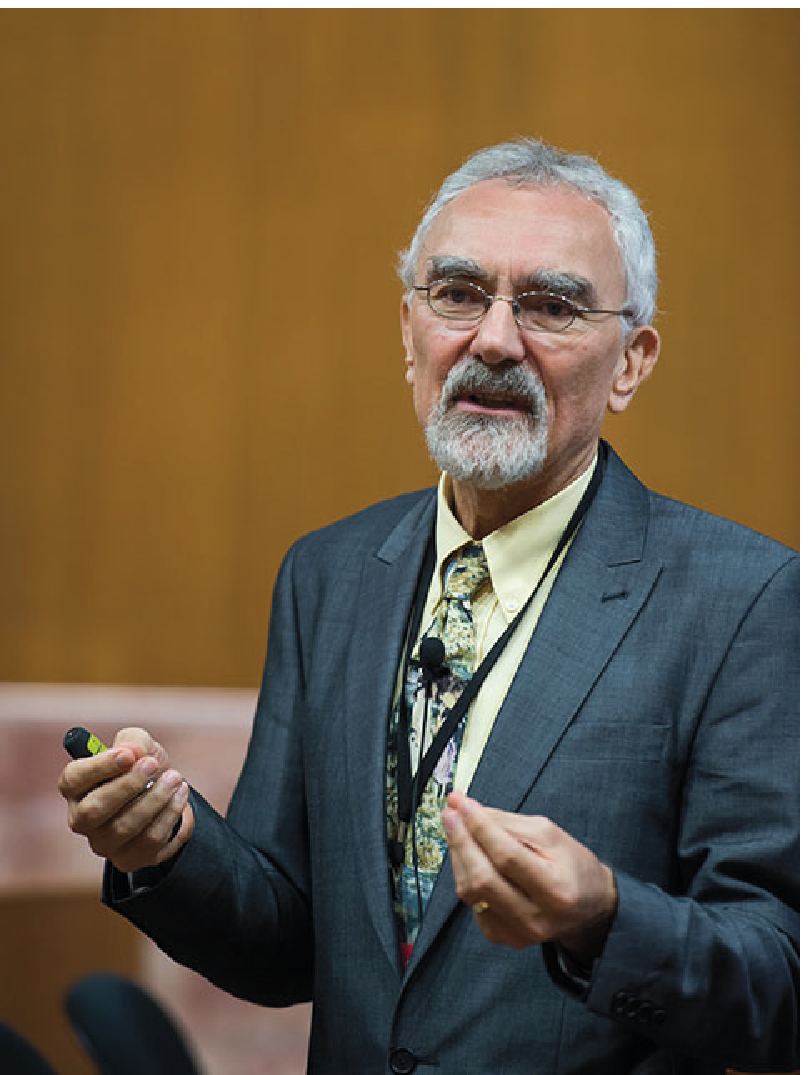}}]
		{Lajos Hanzo}  (http://www-mobile.ecs.soton.ac.uk, https://en.wikipedia.org/wiki/Lajos\_Hanzo) (FIEEE'04) received his Master degree and Doctorate in 1976 and 1983, respectively from the Technical University (TU) of Budapest. He was also awarded the Doctor of Sciences (DSc) degree by the University of Southampton (2004) and Honorary Doctorates by the TU of Budapest (2009) and by the University of Edinburgh (2015).  He is a Foreign Member of the Hungarian Academy of Sciences and a former Editor-in-Chief of the IEEE Press.  He has served several terms as Governor of both IEEE ComSoc and of VTS.  He has published 2000+ contributions at IEEE Xplore, 19 Wiley-IEEE Press books and has helped the fast-track career of 123 PhD students. Over 40 of them are Professors at various stages of their careers in academia and many of them are leading scientists in the wireless industry. He is also a Fellow of the Royal Academy of Engineering (FREng), of the IET and of EURASIP. He is the recipient of the 2022 Eric Sumner Field Award.
\end{IEEEbiography}
\end{document}